%% file: arXivV1.tex
\documentclass[final,twocolumn,5p,times,10pt]{elsarticle}

\usepackage{lineno,hyperref}
\modulolinenumbers[5]

\journal{Automatica}

\biboptions{authoryear}

\usepackage{bm}
\usepackage{amsmath}
\usepackage{amsfonts}
\usepackage{amsthm}
\usepackage{color}
\usepackage{import}
\usepackage{soul}
\usepackage{xcolor}
\usepackage{comment}

\newtheorem{theorem}{\bf{Theorem}}
\newtheorem{lemma}{\bf{Lemma}}

\newtheorem{property}{\bf{Property}}
\newtheorem{remark}{\bf{Remark}}
\newtheorem{assumption}{\bf{Assumption}}
\newtheorem*{problem}{\bf{Problem}}
\newtheorem{claim}{\bf{Claim}}

\definecolor{orange}{RGB}{255,69,0}
\def\ver#1{{\color{black}#1}}
\def\mage#1{{\color{black}#1}}
\def\azul#1{{\color{black}#1}}

\begin{document}

\begin{frontmatter}

\title{Set-Point Tracking MPC with Avoidance Features\tnoteref{mytitlenote}}
\tnotetext[mytitlenote]{\textcopyright 2023. This manuscript version is made available under the CC-BY-NC-ND 4.0 license \url{https://creativecommons.org/licenses/by-nc-nd/4.0/}. This work was supported by the projects INCT InSAC and Universal under the grants CNPq 465755/2014-3 and 426392/2016-7, and by the Brazilian agencies CAPES and FAPEMIG.}

\author[1,3]{Marcelo A. Santos \corref{cor}}
\cortext[cor]{Corresponding author, cidomg32@ufmg.br.} 

\author[2]{Antonio Ferramosca} 

\author[1,3]{Guilherme V. Raffo}

\address[1]{Graduate Program in Electrical Engineering, Federal University of Minas Gerais, Belo Horizonte, MG 31270-901, Brazil}
\address[2]{Department of Management, Information and Production Engineering, University of Bergamo, Via Marconi 5, Dalmine (BG) 24044, Italy}
\address[3]{Department of Electronics Engineering, Federal University of Minas Gerais, Belo Horizonte, MG 31270-901, Brazil}

\begin{abstract}
This work proposes a finite-horizon optimal control strategy to solve the tracking problem while providing avoidance features to the closed-loop system. Inspired by the set-point tracking model predictive control (MPC) framework, the central idea of including artificial variables into the optimal control problem is considered. This approach allows us to add avoidance features into the set-point tracking MPC strategy without losing the properties of an enlarged domain of attraction and feasibility insurances in the face of any changing reference. Besides, the artificial variables are considered together with an avoidance cost functional to establish the basis of the strategy, maintaining the recursive feasibility property in the presence of \ver{a previously unknown} number of regions to be avoided. It is shown that the \ver{closed-loop system} is recursively feasible and \ver{input-to-state-stable} under the mild assumption that the avoidance cost is uniformly bounded over time. Finally, two numerical examples illustrate the controller behavior.

\end{abstract}


\end{frontmatter}

\section{Introduction}

Model predictive control (MPC) is one of the few strategies that allow the control of constrained systems regarding an optimal criterium while ensuring stability and convergence to an equilibrium point. Under certain assumptions, closed-loop stability can be demonstrated for any feasible initial condition in standard \ver{regulating} MPC schemes \citep{Mayne2000}. Often, MPC strategies are designed with the underlying assumption that the desired set-point is feasible, which may not be true since feasibility issues can come from the system dynamics limitations and constraints. Besides, if changing set-points are considered, there is no guarantee that the closed-loop system will \mage{be stable} \citep{Pannocchia2005}. 

The problem of changing set-points has been addressed through different approaches \citep{Mayne2014}, \azul{such as} switching strategies for feasibility recovery \citep{Chisci2003} and command governor-based strategies \citep{Bemporad1997, Garone2017}. \azul{In the command governor framework, a feasible evolution of the system to the reference is obtained based on the inclusion of a low-pass filter of the reference. Inspired by the reference governor ideas, in \citet{Limon2008}, a set-point tracking MPC strategy has been proposed seeking to deal with the problem of loss of recursive feasibility in the presence of changing set-points.} The authors have shown that the so-called tracking MPC has a larger domain of attraction when compared to the ones obtained with \ver{regulating} predictive controllers. In this strategy, the controller is designed to ensure asymptotic convergence for any admissible steady-state reference and, if not the case, to ensure convergence to an admissible steady-state. The control strategy is formulated using artificial state and input variables to describe an artificial steady-state for which asymptotic convergence is guaranteed. In addition, terminal conditions for stability are considered \mage{through} a terminal penalty term in the cost functional and a terminal constraint ensuring that the system reaches a maximum admissible invariant set for tracking. In \citet{Ferramosca2009}, \mage{it was} shown that the controller proposed in \citet{Limon2008} holds the local optimality property present in \ver{regulating} predictive controllers.

In some applications, besides being able to track changing set-points, avoiding specific regions in some known admissible space is an important requirement. For instance, in autonomous navigation, the ability to ensure collision avoidance against obstacles is paramount. \mage{Also,} problems that have inherently non-convex admissible spaces, such as charging Li-ion batteries \citep{Goldar2020}, may benefit from it by going from a non-convex control problem to an equivalent convex one. In the literature, the avoidance problem is often solved through optimal control by considering either the space to be avoided as a modified constraint in an equivalent problem or by adding relaxed avoidance constraints. \ver{Such solutions can be formulated into single-layer frameworks, which, unlike multi-layer strategies, avoid suboptimal solutions, loss of feasibility, and lack of stability guarantees \citep{Limon2012}.}

Among those works considering constraints in an equivalent problem, in \citet{Sasa2021}, an MPC strategy for collision avoidance has been proposed to the regulation problem with deterministic linear systems. The authors \ver{have considered} a strategic-tactical decision-making architecture to obtain a resulting convex MPC for collision avoidance. In \citet{Zhang2021}, the authors have considered avoidance in an $n$-dimensional space by reformulating the collision avoidance constraints as smooth non-convex constraints under the assumption that obstacles can be described as convex sets. In \citet{Thirugnanam2022}, the problem of avoidance between polytopes has been approached using control barrier function constraints to generate dynamically collision-free feasible trajectories. In \citet{Cotorruelo2021}, the authors have extended the set-point tracking MPC framework by incorporating a convexifying homeomorphism into the optimization problem so that non-convex admissible output sets could be handled. If the convexifying homeomorphism exists, the proposed approach \ver{is suitable for applications with non-convex output space.}

Among those works relaxing avoidance constraints, in \citet{Kamel2017}, an avoidance cost \mage{based on} a logistic function has been added to a decentralized nonlinear model predictive control scheme for collision avoidance during multi-agent flights. In \citet{Hermans2018}, the authors have considered nonlinear model predictive control within a penalty method framework designed to satisfy collision-avoidance while calculating the trajectory to be followed by the system. In \citet{Pereira2021}, an ellipsoidal-polytopic representation of obstacles has been incorporated into a nonlinear model predictive controller as relaxed avoidance constraints. Furthermore, in \citet{Sanchez2021}, artificial variables have been used to integrate the obstacle avoidance feature to model predictive control. In the work, obstacles are represented as soft constraints in the optimization problem, and the artificial variables help solving the path-following problem while avoiding obstacles.

Approach the avoidance problem through an equivalent 
\mage{one} that either convexifies the space or changes the constraints to achieve some given properties may require prior knowledge of the regions to be avoided. \mage{Also}, in the presence of changing set-points, feasibility issues must be addressed. When considering \ver{a previously unknown} number of regions to be avoided, the inclusion of penalty functions into the optimization problem allows avoiding online computation of equivalent constraints.  Likewise, the stability analysis can be performed without any assumptions on the convexity of the admissible space, but at the price of possibly invalidating the decreasing property of the value function commonly used to derive stability. A similar problem appears in economic MPC schemes, where the value function is generally non-decreasing as the system approaches the economically optimal steady-state\mage{.} Under this condition, asymptotic stability of the resulting closed-loop system can still be demonstrated. \mage{In} \citet{Diehl2011}\mage{,} a decreasing rotated value function has been designed using dissipativity theory, and in \citet{Alessandretti2017}, it has been shown that any additional cost acting as a disturbance to the standard stabilizing cost presents Input-to-State-Stability (ISS) property as long as it is uniformly bounded over time.

This work differs from the existing literature by approaching the problem in which \ver{a previously unknown} number of regions is considered to be avoided within the system admissible states, while the feasibility issues related to changing set-points are addressed. As in \citet{Limon2008}, the problem of feasibility while tracking changing set-points is solved using artificial variables and an offset cost functional. \ver{Afterwards, the avoidance feature is included in the set-point tracking MPC strategy through a penalty function allowing to work with convex admissible sets even in the presence of a previously unknown number of regions to be avoided. Therefore, with both ideas combined, the resulting control framework is a linear set-point tracking MPC with avoidance features, since, as in \citet{Limon2008}, a linear model is used for prediction. Further, we prove that the proposed set-point tracking MPC with avoidance features have the properties of recursive feasibility and ISS with respect to the avoidance cost functional.}

The contributions of this work are threefold: \mage{i) a novel linear set-point tracking MPC strategy with avoidance features; ii) proof of recursive feasibility for changing targets and \ver{previously unknown} non-feasible output regions \ver{to be avoided within the system admissible states}; and iii) demonstration of ISS property with respect to the avoidance cost.}

\mage{In the remainder of this paper:} Section 2 describes the problem addressed; Section 3 presents the control design and demonstrates recursive feasibility and \ver{ISS}; Section 4 analyses the proposed controller properties; \mage{Section 5 provides two numerical examples; finally, Section 6 concludes the work.}

\subsection*{\textbf{Notation and definitions}}

The set $\mathbb{I}_{0:N}$ denotes the set of integers $\{0, 1, \cdots, N\}$. A matrix $O_{n,m} \in \mathbb{R}^{n\times m}$ denotes a matrix of zeros, and $I_n \in \mathbb{R}^{n\times n}$ denotes the identity matrix. A positive definite symmetric matrix $P$ is denoted as $P > 0$, and $\|x\|_P = \sqrt{x'Px}$ denotes the weighted Euclidean norm of $x$ with $'$ being the transpose operator. Consider $a \in \mathbb{R}^{n_a}$ and $b \in \mathbb{R}^{n_b}$, for a set $\Gamma \subset \mathbb{R}^{n_a+n_b}$, the projection operation is defined as $\textrm{Proj}_{a}(\Gamma) = \{a \in \mathbb{R}^{n_a} : \exists b \in \mathbb{R}^{n_b}, (a,b) \in \Gamma\}$. Given a set $X\subset\mathbb{R}^n$ and a variable $\lambda \in \mathbb{R}$, the set $\lambda X \subset \mathbb{R}^n$ is defined as $\lambda X = \{\lambda x: x \in X\}$. A bold lowercase variable $\bm{u}$ denotes a sequence of values of a signal $(u(0), u(1), \cdots, u(N-1))$, with $u(i)$ being the $i$-th element and $N$ being the length of the sequence deduced by the context. A parameter-dependent signal is denoted by $\bm{u}(a)$, and its $i$-th element is $u(i;a)$. The identity function from $\mathbb{R}$ onto $\mathbb{R}$ is denoted as $id$, and  $\gamma_1 \circ \gamma_2$ denotes the composition of two functions $\gamma_1$ and $\gamma_2$. A diagonal matrix $M \in \mathbb{R}^{n\times n}$ is denoted as $M = \mathrm{diag}\{m_1, \cdots, m_n\}$, and $\textrm{max}\{a_1, \cdots, a_n\}$ denotes the maximum operator. 

\section{Problem Description}

Consider a linear time-invariant dynamical system of the form
\begin{equation}
	\label{eq:nominalmodel}
	\begin{array}{rcl}
		x(k+1) & = & A x(k) + B u(k),\\
		y(k) & = & C x(k) + D u(k),
	\end{array}
\end{equation}
with $x(k) \in \mathbb{R}^n$, $u(k) \in \mathbb{R}^m$, and $y(k) \in \mathbb{R}^p$ being, respectively, the state, input, and output vectors. 
The solution of the system for a given sequence of control inputs $\bm{u}$ and initial state $x$ is denoted as $x(j) = \phi(j; x, \bm{u})$, $j \in \mathbb{I}_{\geq 0}$, where $x = \phi(0; x, \bm{u})$.
\begin{assumption}\label{assump1}
	The dynamical system \eqref{eq:nominalmodel} is controllable, observable, and the states are available at each sampling time.
\end{assumption}
The evolution of the system must be such that the constraint  
\begin{equation}
	\label{eq:constraints}
	(x(k), u(k)) \in Z
\end{equation}
holds for all $k \geq 0$, defining the sets of admissible states and inputs as $X = \textrm{Proj}_x(Z)$ and $U = \textrm{Proj}_u(Z)$, respectively. \ver{In addition}, there exists an invertible linear map $f: Z \mapsto Y$ defining the set of admissible output $Y$.
\begin{assumption}\label{assump3}
	The set $Z \subset \mathbb{R}^{n+m}$ is a compact convex polyhedron containing the origin in its interior.
\end{assumption}

If a finite \ver{previously unknown number $N_o$ of non-feasible} output regions $\azul{O_i}$ strictly contained in $Y$ are considered, the admissible output set might become \mage{the non-convex set} 
\begin{equation}
	\label{eq:nonconvecoutput}
	\tilde{Y} = Y - \bigcup\nolimits_{i=1}^{N_o} \azul{O_i}.
\end{equation}
\ver{Thus, assuming that the inverse map $f^{-1}: \tilde{Y} \mapsto \tilde{Z}$ exists, the evolution of the system must be such that the constraint}
\begin{equation}
	\label{eq:nonconveconstraints}
	(x(k), u(k)) \in \tilde{Z}
\end{equation}
is satisfied for all $k \geq 0$, where $\tilde{Z}$ may be a non-convex set that does not fulfill Assumption \ref{assump3}.

Set-point tracking MPC aims to make the error between the target output and the actual output tend to zero. Without the presence of non-feasible output regions $\azul{O_i}$, for asymptotic stabilization, a target output $y_t$ must be a steady-output associated with an admissible equilibrium point $\mage{(x_s, u_s) \in Z}$. If this condition is satisfied, the target output is said to be reachable; otherwise, the tracking control problem fails since it is not possible to stabilize the system in the provided output \citep{Limon2008}. Therefore, any target output must satisfy
\begin{equation}
	\label{eq:equilibriumcondition}
	\begin{bmatrix}
		A - I_n & B\\
		C	   & D
	\end{bmatrix}
	\begin{bmatrix}
		x_s \\ u_s 
	\end{bmatrix}
	= 
	\begin{bmatrix}
		O_{n,1} \\ y_t
	\end{bmatrix}.
\end{equation}

For any given target output $y_t$, there exists an associated equilibrium point $(x_s, u_s)$ if and only if \citep{Rawlings2009}
\begin{equation}
	\label{eq:steadycondition}
	\textrm{rank}
	\left(
	\begin{bmatrix}
		(A-I_n) & B \\
		C      & D
	\end{bmatrix}
	\right)
	= n + p.
\end{equation}
For a square system, $p = m$, if condition \eqref{eq:steadycondition} is satisfied, every $y_t$ can be tracked and there is a unique equilibrium point $(x_s, u_s)$ associated to it. For a flat system, $p < m$, if condition \eqref{eq:steadycondition} holds, every $y_t$ can be tracked and there is an infinite number of equilibrium points $(x_s, u_s)$ whose output is $y_t$. If the system is thin, \ver{$ p > m$, or the condition \eqref{eq:steadycondition} does not hold}, $y_t$ can be partially tracked or steered to a target zone \citep{Ferramosca2010}.

Considering that condition \eqref{eq:steadycondition} is satisfied, it is possible to define the set of joint steady-states and inputs, $Z_s$, and the set of reachable outputs, $Y_r$, respectively, as
\begin{align}
	Z_s &= \{(x_s, u_s) : x_s = Ax_s + Bu_s, (x_s, u_s) \in Z\}, \label{eq:zreachableset} \\
	Y_r &= \{y_t : y_t = Cx_s + Du_s, (x_s, u_s) \in Z_s\}. \label{eq:outputreachableset}
\end{align}

Two main sources of feasibility and stability issues are present when handling set-point tracking MPC with avoidance features. First, since the feasibility region for the closed-loop system is reference-dependent, unknown variations on the target outputs may compromise recursive feasibility and asymptotic stability, leading the controller to fail tracking the reference \citep{Pannocchia2005}. Second, the existence of \ver{previously unknown} non-feasible regions in the known admissible output space might make the target output unfeasible ($y_t \notin \tilde{Y}$) or unreachable in the obstructed space (there is no evolution of the system output towards $y_t$ that fulfill $y(k) \in \tilde{Y}, \forall k \geq 0$). Therefore, within this context, the following problem is posed:
\begin{problem}
Design an MPC law $\kappa_N^o(x(k), y_t, \azul{O_i})$ to track any prior reachable target output $y_t \in Y_r$ ensuring that the evolution of the system output lies outside any non-feasible output region $\azul{O_i}$. \ver{Also, by considering that a global solution to the problem can be obtained, if the target is feasible and reachable in the obstructed space ($y_t \in Y_r \cap \tilde{Y}$), and there is no non-feasible region in the neighborhood of the target, the tracking error must tend to zero asymptotically. Otherwise, the system output must converge to a bounded set around a reachable steady-output $y_s \in Y_r \subset \tilde{Y}$ that minimizes a given performance index.}
\end{problem}

\section{Control Design}

The controller proposed in this section is designed to ensure \ver{Input-to-State Stability} in the Lyapunov sense for any reachable target in the obstructed space while avoiding any non-feasible output region $\azul{O_i}$. \mage{We} consider the system \eqref{eq:nominalmodel} subjected \mage{to} the constraint \eqref{eq:constraints}. Besides, to avoid loss of controllability related to active constraints \citep{Rao1999}, we remove from $Y_r$ those reachable targets that are associated with equilibrium points lying at active constraints. \mage{Then}, the condition $(x_s, u_s) \in Z_s$ of \eqref{eq:outputreachableset} is replaced by $(x_s,u_s) \in \lambda Z_s$ with $\lambda \in (0,1)$ \ver{and possibly very close to 1}.

Avoidance features can be obtained enforcing $y(k) \in \tilde{Y}$ for all $k>0$, implying that the closed-loop system satisfies \eqref{eq:nonconveconstraints}. However, since $\tilde{Z}$ is possibly a non-convex set priory unknown, enforcing \eqref{eq:nonconveconstraints} directly may be impractical from the optimization problem point-of-view. \ver{Besides, there is no guarantee that the inverse map $f^{-1}: \tilde{Y} \mapsto \tilde{Z}$, required to obtain \eqref{eq:nonconveconstraints}, exists. A possible solution to work around those issues considers an equivalent strictly convex optimization problem that constrains the closed-loop system to the known admissible convex set $Z$ by enforcing \eqref{eq:constraints} and that handles the non-feasible regions $\azul{O_i}$ through penalties functions.} Following this procedure, set-point tracking MPC with avoidance features can be obtained extending the formulation proposed in \citet{Limon2008} by considering an avoidance cost functional in addition to the offset functional.

Let $y_a$ be an artificial steady-output, which is an extra decision variable in the optimal control problem to avoid issues related to the loss of feasibility. Moreover, let $(x_a, u_a)$ be an artificial equilibrium point associated with $y_a$.  

\begin{assumption}
	\label{assump4}
	\ver{Any output non-feasible set $\azul{O_i}$ is available at each sampling time either by measurement and estimation or, if available, by previous knowledge on the sets. Also, they are considered constant throughout the prediction horizon.}
\end{assumption}

\begin{remark}
	\ver{Modeling, measurement, and estimation errors can be accounted for considering an enclosure for each set $\azul{O_i}$, such as  $\azul{\sigma_i O_i}$ with $\azul{\sigma_i > 1}$.}
\end{remark}
 
\azul{Given a value function $V(y)$ and the constraints $y \notin O_i$, for all $i$, with the constraint set being represented as $O_i = \{y: g_j(y) \leq 0, j \in \mathbb{I}_{1:q_i}\}$, the optimization problem} 
\begin{align}
	\mathrm{minimize} & \quad V(y) & \text{subject to} & \quad \azul{y \notin O_i, \forall i}, 
\end{align}
can be rewritten as the unconstrained problem 
\begin{align}
	\mathrm{minimize} & \quad V(y) + \azul{\sum_i\mu_i F(y,O_i)},
\end{align}
where $\azul{\mu_i}$ is a positive constant, and $F(y,\azul{O_i})$ is a continuous function such that $F(y,\azul{O_i}) \geq 0$ if $\azul{y \in O_i} $\ver{,} and $F(y,\azul{O_i}) = 0$ otherwise \citep{Luenberger2008}. A general class of penalty functions is 
\begin{equation}
	\label{eq:penalty}
	F(y,\azul{O_i}) = \sum\nolimits_{j = 1}^{\azul{q_i}} (\mathrm{max}\{0,\azul{g_j(y)}\})^{\epsilon},
\end{equation}
for some $\epsilon > 0$. As $\azul{\mu_i} \rightarrow \infty$, the solution of the penalty problem converges to the solution of the constrained problem. Besides, if $\epsilon = 1$, exact penalization can be obtained if $\azul{\mu_i}$ is chosen to be greater than the biggest corresponding Lagrange multiplier \citep{Luenberger2008,Ferramosca2011}. \azul{Notice that the definition of $\mu_i$ allows us to give different weights for different non-feasible regions $O_i$.}

Following the presented penalty method, the proposed controller is based on the solution at each sampling time of an optimal control problem having as parameters $(x, y_t, \azul{O_i})$ and as decision variables $(\bm{u}, x_a, u_a)$. The cost functional is composed of three terms: i) a dynamic term, which is a combination of a stage cost with respect to the artificial steady-state and input $(x_a, u_a)$ and a terminal cost; ii) a stationary term, which is the offset cost functional penalizing the deviation of the artificial steady-output $y_a$ to the target output $y_t$; and iii) another stationary term, which is the avoidance cost functional penalizing the artificial steady-output $y_a$ and the system predicted output $\bm{y}$. \mage{Considering a horizon length $N \in \mathbb{I}_{> 0}$, it can be defined}
\begin{align}
	\label{eq:costfunction}
	V_{N}(&x, y_t, \azul{O_i}; \bm{u},  x_{a}, u_{a}) = \sum\limits_{j=0}^{N-1}\|x(j) -  x_a\|_{Q}^2 + \| u(j) - u_a \|^2_{R} +  \nonumber\\
	& \|x(N) -  x_a\|_{P}^2 + V_{of}(y_a, y_t) + \sum\limits_{i = 1}^{N_o}\Big[\azul{\mu_i} F(y_a,\azul{O_i}) + \nonumber \\
	& \sum\limits_{j=0}^{N}\azul{\mu_i} F(y(j),\azul{O_i})\Big]. 
\end{align}

The following assumptions from the tracking MPC literature are sufficient conditions to ensure asymptotic stability for the closed-loop system \ver{without non-feasible regions} $\azul{O_i}$ \citep{Limon2008, Ferramosca2009}.
\begin{assumption}\label{assump5}
		Let $R \in \mathbb{R}^{m \times m}$ be a positive definite matrix and $Q \in \mathbb{R}^{n\times n}$ a positive semidefinite matrix such that the pair $(Q^{1/2}, A)$ is observable.
\end{assumption}
\begin{assumption}\label{assump5a}
		Let $K \in \mathbb{R}^{m\times n}$ be a stabilizing control gain such that the matrix $A_{K} = A + BK$ is Schur.
\end{assumption}
\begin{assumption}\label{assump5b}
		Let $P \in \mathbb{R}^{n \times n }$ be a positive definite matrix, solution of the Lyapunov equation \mage{${P = A_{K}'PA_{K}^{} + Q + K'RK}$}.
\end{assumption}
\begin{assumption}\label{assump5c}
		Let $\Omega_t^a \subseteq \mathbb{R}^{n+n+m}$ be an admissible polyhedral invariant set for tracking for \eqref{eq:nominalmodel} subject to \eqref{eq:constraints}, \mage{for a given gain $K$. That is, $\forall (x, x_a, u_a) \in \Omega_t^a$, it holds that $(x, K(x - x_a) + u_a) \in Z$, $(x_a, u_a) \in \lambda Z_s$, and $(Ax + B(K(x- x_a) + u_a), x_a, u_a) \in \Omega_t^a$.}
\end{assumption}
\begin{assumption}\label{assump5d}
		Let the offset cost $V_{of}\left(\cdot\right) : \ver{\mathbb{R}^{2p}} \mapsto \mathbb{R}_{\geq 0}$ be a continuous, convex, and positive definite function with $V_{of}\left(0, 0\right) = 0$ for $k=0$, such that \mage{$\mathrm{arg} ~ \underset{y_a \in Y_r}{\mathrm{min}} ~V_{of}\left(y_a, y_t\right)$} is unique \ver{for any $y_t$}.
\end{assumption}

\ver{Defining $\Omega_t = \mathrm{Proj}_x(\Omega_t^a)$, the feasible region $X_N(\Omega_t)$ is defined as the $N$-steps controllable set to $\Omega_t$. Notice that $X_N(\Omega_t)$ is by definition the domain of attraction for the proposed controller.} Furthermore, to provide the controller with avoidance features and to later derive the ISS property with respect to the avoidance cost, consider the following assumption.
\begin{assumption}
	\label{assump6}
	Let $V_{av}\left(\cdot\right) : \mathbb{R}^p \mapsto \mathbb{R}_{\geq 0}$ be the continuous function $V_{av}\left(\cdot\right) = \sum_{i=1}^{N_o}\big[\azul{\mu_i} F(y_a,\azul{O_i}) + \sum_{j=0}^{N}\azul{\mu_i} F(y(j),\azul{O_i})\big]$. Moreover, \ver{let the bound of the avoidance function be defined as $\azul{S} = \mathrm{sup}(V_{av}(\cdot ))$, such that} $V_{av}(\cdot) \rightarrow \azul{S}$ if $y_a\notin \tilde{Y}$ or $y(j)\notin \tilde{Y}$ for any $j \in \mathbb{I}_{0:N}$, with $V_{av}(\cdot) = 0$ whenever $y_a\in \tilde{Y}$ and $y(j)\in \tilde{Y}$ for all 
 $j \in \mathbb{I}_{0:N}$. 
\end{assumption}

The controller is derived from the solution of the optimization problem $P_N^O(x, y_t,\azul{O_i})$ given by
\begin{subequations} 
	\label{eq:mpc}
	\begin{align}
		V_{N}^{O}(x,y_t, \azul{O_i}) =  \underset{\bm{u}, x_{a}, u_{a}}{\mathrm{min}} & V_{N}(x, y_t, \azul{O_i} ; \bm{u},  x_{a}, u_{a}) \nonumber \\
		\text { s.t. } & x(0)=x, \label{eq:const1} \\
		&x(j+1) = Ax(j) + Bu(j), \label{eq:const2} \\
		& \ver{y(j) = Cx(j) + Du(j),} \label{eq:const2_1} \\
		& (x(j), u(j)) \in Z, ~ j \in \mathbb{I}_{0:N-1}, \label{eq:const3}\\
		&{y_a = Cx_{a} + Du_a}, \label{eq:const4} \\
		&(x(N), x_a, u_a) \in \Omega_t^a, \label{eq:const5}
	\end{align}
\end{subequations} 
with constraints \eqref{eq:const1}-\eqref{eq:const3} subjecting the predicted trajectory to the system dynamics and constraints, and with constraints \eqref{eq:const4} and \eqref{eq:const5}, respectively, defining the artificial steady-output related to an artificial equilibrium and enforcing the terminal state to be in a region where the system can be stabilized by a local control law $u = K(x - x_a) + u_a$. Notice that the constraints of the problem $P_N^O(x, y_t, \azul{O_i})$ do not depend on $y_t$, making it feasible for any changing set-point. \ver{Additionally, the resulting optimization problem has a known convex output space because the inclusion of the previously unknown non-feasible regions $\azul{O_i}$ as penalties allowed the use of the known admissible set $Z$ in constraint \eqref{eq:const3}. Furthermore, the penalty approach makes the set $\Omega_t^a$ time-invariant, which allows it to be obtained through offline computation.}

Considering the receding policy of MPC controllers and \mage{that the problem} \eqref{eq:mpc} is solved at each sampling time based on the current knowledge of the optimization parameters, the optimal control law is given by \mage{$\kappa_N^o(x, y_t, \azul{O_i}) = u^O(0;x, y_t, \azul{O_i})$.}

\begin{theorem}(Asymptotic stability \citep[Theorem 1]{Ferramosca2009})
	\label{theoremAntonio}
	Consider that Assumptions \ref{assump1}, \ref{assump3}, and \ref{assump5} to \ref{assump5d} hold for \eqref{eq:nominalmodel} constrained by \eqref{eq:constraints} without the presence of non-feasible output regions $\azul{O_i}$. For a given target $y_t$ and for any feasible initial state $x \in X_N(\Omega_t)$, the closed-loop system with $\kappa_N^o(x, y_t)$ is stable, fulfills the constraints throughout the time and, besides
	\begin{itemize}
		\item[(i)] If $y_t \in Y_r$, the closed-loop system asymptotically converges to $y_t$.
		\item[(ii)] If $y_t \notin Y_r$, the closed-loop system asymptotically converges to a \ver{reachable steady-output that minimizes $V_{of}\left(y_a, y_t\right)$}.
	\end{itemize}
\end{theorem}

\begin{theorem}(ISS-based avoidance) 
	\label{theorem}
	Consider that Assumptions \ref{assump1} to \ref{assump6} hold, then the closed-loop system with $\kappa_N^o(x, y_t, \azul{O_i})$ is ISS with respect to the avoidance cost $V_{av}(\cdot)$, i.e., there is a $\mathcal{KL}$-function $\beta(\cdot)$ and a $\mathcal{K}$-function $\gamma(\cdot)$ such that for any feasible initial state $x(0) \in X_N(\Omega_t)$, \ver{steady-state $x_s \in Z_s$}, and bound $\azul{S}$, the solution $\phi(k; x(0), \bm{u})$ exists and satisfies \mage{for all $k \in \mathbb{I}_{> 0}$}
	\begin{equation}
		\|\phi(k; x(0), \bm{u}) - \ver{x_s}\| \leq \beta(\|x(0) - \ver{x_s}\|,k) + \gamma(\azul{S}).
	\end{equation}	
\end{theorem}

\ver{Theorem \ref{theorem} can be interpreted as follows. In presence of non-feasible output regions, the avoidance cost acts as disturbance and only ISS can be ensured. Therefore, the closed-loop system converges to a bounded set around a steady-state, either desired or feasible. \mage{Besides, depending on the penalties obtained, only local convergence might be achieved.} In this context, asymptotic stability in the terms of Theorem \ref{theoremAntonio} is only recovered when the avoidance cost goes to zero.}

To demonstrate Theorem \ref{theorem}, first, we prove that the controlled system is recursively feasible. Afterward, as proposed in \citet{Alessandretti2017}, a shifted value function is defined to account for the effect of the avoidance cost. Then, upper and lower bounds are obtained, as well as a bound on the shifted value function decrease. Finally, inspired by \citet{Alessandretti2017} and  following the procedure presented in \citet{Jiang2001}, it is shown that the closed-loop system is ISS with respect to the bound $\azul{S}$ and, consequently, to the avoidance cost functional. Proofs of the following lemmas can be found in the Appendix.

\begin{lemma}(Steady condition convergence)
	\label{lemma:Limon2018}
	\ver{Consider that Assumptions \ref{assump1} to \ref{assump6} hold for the system \eqref{eq:nominalmodel} constrained by \eqref{eq:constraints}. For any feasible initial state $x \in X_N(\Omega_t)$, target $y_t$, and bound $\azul{S}$, let the optimal solution to $P_N^O(x, y_t,\azul{O_i})$ be such that $x = x_a^O$, $u = u_a^O$, and $y = y_a^O$. Moreover, let $(x_s, u_s, y_s)$ be the optimal triplet satisfying \eqref{eq:equilibriumcondition}, such that function $V_{of}(y_a, y_t) + V_{av}(\bm{y}, y_a,\azul{O_i})$ is minimized. Then, $x = x_s$, $u = u_s$, and $y = y_s$.}
\end{lemma}
\begin{lemma}(Artificial error boundedness)
	\label{lemma:error}
	Consider that the Assumptions \ref{assump1} to \ref{assump6} hold. \ver{Let $x_s$ be the optimal steady-state associated to the optimal target $y_s$, such that function $V_{N}(x,y_t,\azul{O_i})$ is minimized.} For all $x \in X_N(\Omega_t)$ and $x_a^O \in \textrm{Proj}_x(Z_s)$, define the function $e(x) = x - x_a^O$. Then, there exists a $\mathcal{K}$-function $\alpha_e(\cdot)$ such that \mage{$\|e(x)\| \geq \alpha_e(\|x-x_s\|)$}.
\end{lemma}
\begin{lemma}(Recursive feasibility)
	\label{lemma1}
	Consider that Assumptions \ref{assump1} to \ref{assump5d} hold, then the closed-loop system with $\kappa_N^o(x, y_t, \azul{O_i})$ is recursively feasible for any feasible state $x \in X_N(\Omega_t)$.
\end{lemma}

\mage{Consider as a Lyapunov candidate for the problem $P_N^O(x, y_t, \azul{O_i})$ the shifted value function defined as $V_s(x, y_t, \azul{O_i}) = V_N(x, y_t, \azul{O_i}) - \azul{S}$.}

\begin{lemma}(Upper bound)
	\label{lemma2}
	Consider that Assumptions \ref{assump1} to \ref{assump6} hold \mage{and let $\alpha_c(\cdot)$ be a $\mathcal{K}_{\infty}$-function}, then the shifted value function $V_s(\cdot)$ satisfies \mage{$V_s(x, y_t,\azul{O_i}) \leq \alpha_c(\|x - x_s\|)$.}
\end{lemma}

\begin{lemma} (Lower bound)
	\label{lemma3}
	Consider that Assumptions \ref{assump1} to \ref{assump6} hold \mage{and let $\alpha_b(\cdot)$ be a $\mathcal{K}_{\infty}$-function}, then the shifted value function $V_s(\cdot)$ satisfies \mage{$V_s(x, y_t, \azul{O_i}) \geq \alpha_b(\|x - x_s\|) - \azul{S}$}.
\end{lemma}

\begin{lemma}(Decreasing property)
	\label{lemma4}
	Consider that Assumptions \ref{assump1} to \ref{assump6} hold \mage{and let $\alpha(\cdot)$ be a $\mathcal{K}_{\infty}$-function}, then the shifted value  function $V_s(\cdot)$ satisfies \mage{$V_s^O(x(k+1), y_t, \azul{O_i}) - V_s^O(x(k), y_t, \azul{O_i}) \leq -\alpha(\|x - x_s\|) + \azul{S}$}.
\end{lemma}

\begin{lemma}(ISS bound)
	\label{lemma5}
	Consider that Assumptions \ref{assump1} to \ref{assump6} hold, then there exists a $\mathcal{KL}$-function $\hat{\beta}$ such that the shifted function $V_s(\cdot)$ satisfies \mage{$V_s^O(x(k),y_t, \azul{O_i}) \leq \mathrm{max}\{\hat{\beta}(V_s^O(x(0),y_t, \azul{O_i}), k), \hat{\gamma}(\azul{S})\}$, where $\hat{\gamma}(r) = \hat{\alpha}^{-1}\circ\rho^{-1}(r)$} for all $k \in \mathbb{I}_{\geq 0}$, with $\rho(\cdot)$ being a $\mathcal{K}_{\infty}$-function such that $(id - \rho)(\cdot)$ is a $\mathcal{K}_{\infty}$-function, and with $\hat{\alpha}(\cdot)$ being a $\mathcal{K}_{\infty}$-function such that $\hat{\alpha}(r) \leq \alpha_b\circ\alpha_c^{-1}(r)$, for all $r\geq0$, and with $(id - \hat{\alpha})(\cdot)$ being a $\mathcal{K}$-function.
\end{lemma}

\begin{proof}(\textbf{Theorem 2})
	From Lemma \ref{lemma3} and Lemma \ref{lemma5}, and knowing that $\mathrm{max}\{a,b\}\leq a+b\leq\mathrm{max}\{2a,2b\}$, 
	\begin{align}
	\alpha_b(\|x - x_s\|) & \leq   \mathrm{max}\{\hat{\beta}(V_s^O(x(0), y_t, \azul{O_i}),k), \hat{\gamma}(\azul{S})\} + \azul{S}\nonumber\\
	& \mage{\leq  \mathrm{max}\{2\hat{\beta}(V_s^O(x(0), y_t, \azul{O_i}),k), 2\hat{\gamma}(\azul{S}) , 2\azul{S}\}},
	\end{align}
	and by the monotonicity of $\alpha_b^{-1}(\cdot)$
	\begin{align}
	& \|x-x_s\| = \mage{\alpha_b^{-1}\circ\mathrm{max}\{ 2\hat{\beta}(\alpha_c(\|x(k_0)\ver{-x_s}}\|),k), 2\hat{\gamma}(\azul{S}) , 2\azul{S}\} \nonumber \\
	& \leq  \alpha_b^{-1}\circ2\hat{\beta}(\alpha_c(\|x(k_0)\ver{-x_s}\|),k)) + \alpha_b^{-1}\circ2\hat{\gamma}(\azul{S}) + \alpha_b^{-1}(2\azul{S})\mage{.} 
	\end{align}
	\mage{Moreover, let} $\beta(\cdot)$ and $\gamma(\cdot)$ \mage{be} class-$\mathcal{KL}$ and class-$\mathcal{K}_\infty$ functions, respectively, defined as \mage{$\beta(r_1,s) = \alpha_b^{-1}\circ2\hat{\beta}(\alpha_c(r_1,s))$ and $\gamma(r_2) = \alpha_b^{-1}\circ2\hat{\gamma}(r_2) + \alpha_b^{-1}(2r_2)$}.
	
    \ver{Based on Lemmas \ref{lemma2} to \ref{lemma4}, $V_s(x, y_t, \azul{O_i})$ is a ISS-Lyapunov function for \eqref{eq:nominalmodel} with bounds $\hat{\beta}(\cdot)$ and $\hat{\gamma}(\cdot)$. Then, from \citet[Lemma 3.5]{Jiang2001}, the system is ISS, i.e., $\|\phi(k; x(0), \bm{u})-x_s\| \leq \beta(\|x(0)-x_s\|,k) + \gamma(\azul{S})$ for all $k \in \mathbb{I}_{> 0}$, which concludes the proof.}
\end{proof}

\begin{remark} (Terminal Equality Constraint)
	\label{remark1}
Considering a terminal equality constraint is a practical way to implement the proposed controller. For that, let $(x_a, u_a) \in \lambda Z_s$, $x(N) = x_a$, and $P = O_{n,n}$. Following the same arguments presented before, it can be proved that the results of Theorem \ref{theorem} and Lemma \ref{lemma1} still hold under the mild assumption that the N-controllability matrix of the system, $Co_N = [A^{N-1}B ~ \cdots ~ AB ~ B]$, has full rank. 
\end{remark}

\section{Controller Properties}
\ver{In addition to} the stability guarantees proved before, some properties of the set-point tracking MPC strategy still hold after the inclusion of the avoidance penalty cost.

\begin{property}[\mage{Stability under changing references}]
Theorems \ref{theoremAntonio} and \ref{theorem} show that the closed-loop system with $\kappa_N^o(x, y_t, \azul{O_i})$ is \ver{ISS with respect to the avoidance cost and asymptotic stable for a given target $y_t$ if the avoidance cost tends to zero, i.e., if there are no non-feasible output regions $\azul{O_i}$.} Besides, since the constraints of the problem $P_N^O(x, y_t, \azul{O_i})$ do not depend on $y_t$, the closed-loop system is feasible for any changing target. Thus, even in the presence of significant changes in $y_t$, the controller is still well-posed \ver{and recursive feasibility and input-to-state stability are not lost.}
\end{property}

\begin{property}[\mage{Unreachable references}]
In the case of unreachable references, either due to the system dynamics and constraints or due to non-feasible output regions $\azul{O_i}$, i.e., $y_t \notin Y_r$, $y_t \notin \tilde{Y}$, or $y(k) \notin \tilde{Y}$ for some $k\geq0$, the controller will steer the system to a reachable steady-output $y_s\in Y_r \subset \tilde{Y}$ such that \mage{$y_s = \mathrm{arg} \underset{~y_a \in Y_r}{\mathrm{min}} V_{of}(y_a, y_t) +  V_{av}(\bm{y}, y_a, \azul{O_i})$.} The same behavior occurs in the case of local solutions to the optimization problem.
\end{property}

\begin{property}[\mage{Enlarged domain of attraction}]
Since the designed terminal set $\Omega_{t}$ in the worst case will be equal to the maximal admissible invariant set of a standard MPC \citep{Mayne2000}, the domain of attraction $X_N(\Omega_t)$ is said to be enlarged \mage{(see \citet{Limon2008} for further details)}. 
\end{property}

\begin{property}[\mage{Avoidance guarantees}]
Avoidance can be ensured only if at each sampling period the artificial and the predicted output sequence lie outside any non-feasible output regions $\azul{O_i}$, i.e., \azul{$y_a \notin O_i$ and $\bm{y} \notin O_i$}. These constraints, and consequently avoidance, can be enforced exactly through the penalty function \eqref{eq:penalty} if $\azul{\mu_i} \rightarrow \infty$ or if the penalization is considered with $\epsilon = 1$ and $\azul{\mu_i}$ greater than the biggest corresponding Lagrange multiplier \citep{Luenberger2008,Ferramosca2011}. However, since exact penalization often results in ill-posed optimization problems and $\azul{\mu_i} \rightarrow \infty$ violates Assumption \ref{assump6}, non-exact penalization may be chosen together with a definition of safety regions around the non-feasible regions $\azul{O_i}$ to ensure avoidance while keeping the avoidance cost superiorly bounded, e.g., by considering \ver{$\azul{\sigma_i O_i}$ with $\azul{\sigma_i > 1}$. Furthermore, this cautious approach of considering an enclosure of $\azul{O_i}$ can be exploited to mitigate uncertainties, such as measurement, estimation, modeling, and linearization errors.}
\end{property}

\section{Examples}

This section presents \mage{two simulated examples} obtained to corroborate the effectiveness of the proposed MPC strategy to provide set-point tracking control with avoidance features. The first one considers a ball-on-plate system with a known non-convex admissible output set. The second example considers a UAV navigating in a cluttered environment with \ver{previously unknown} obstacles. In both cases, the simulations are performed with MATLAB\textsuperscript{\textregistered} using the CasADI Toolbox \citep{Andersson2019} with the IPOPT solver \citep{Wchter2005}.

\subsection{Ball-on-plate with non-convex plate}

Consider the ball-on-plate system with non-convex admissible output set proposed in \citet{Cotorruelo2021}. \mage{Based on a reference frame rigidly attached to the center of the plate, let $p_1$ and $p_2$ be the position of the ball and $\theta_1$ and $\theta_2$ be the angle of the plate, both along the reference frame axes.} Thus, the mechanical system can be modeled for simulation purposes as
\begin{equation}
	\label{eq:ballandplatemodel}
	\begin{array}{rcl}
		\ddot{p}_1 = \frac{m}{m+I_b/r^2}(p_1\dot{\theta}_1^2 + p_2\dot{\theta}_1\dot{\theta}_2 + g\sin\theta_1), &  & \mage{\ddot \theta_1 = a_1,}\\
		\ddot{p}_2 = \frac{m}{m+I_b/r^2}(p_2\dot{\theta}_2^2 + p_1\dot{\theta}_1\dot{\theta}_2 + g\sin\theta_2), &  & \mage{\ddot \theta_2 = a_2,} 
	\end{array}
\end{equation}
with the parameters $m = 0.05$ Kg, $r = 0.01$ m, and $I_b = 2.5\cdot 10^{-6}$ Kg$\cdot$m$^2$ being, respectively, the ball mass, radius, and inertia moment. Moreover, $g = 9.81$ m/s$^2$ is the gravitational acceleration.

From \eqref{eq:ballandplatemodel} and considering the system actuated through the desired angular acceleration of the plate, the state and input vectors can be defined as $x = [p_1 ~ p_2 ~ \theta_1 ~ \theta_2 ~ \dot{p}_1 ~ \dot{p}_2 ~ \dot{\theta}_1 ~ \dot{\theta}_2]'$ and $u = [a_1 ~ a_2]'$. Further, the output of the system is $y=[p_1 ~ p_2]'$. The system input is constrained by $\|u\|_{\infty}\leq0.2$, and the admissible output space is constrained by the ellipsoids \mage{$\azul{Y_1} = \{y: (y-y_{c_1})'E_1(y - y_{c_1}) \leq 1\}$} and \mage{$\azul{Y_2} = \{y: (y-y_{c_2})'E_2(y - y_{c_2}) \leq 1\}$}, 
with \mage{$E_1 \!=\! 
	\begin{bmatrix}
		16 \!& \! 0 \\ 0 \! & \! 0.5
	\end{bmatrix}$, 
	$E_2 \!=\!
	\begin{bmatrix}
		5.8551 \!& \!7.3707 \\ 7.3707 \!&\! 10.6449
	\end{bmatrix}$, 	
	$y_{c_1}\! =\! y_{c_2} = 
	\begin{bmatrix}
		0\! & \!0
	\end{bmatrix}'$.}

Following the proposed methodology, the admissible output set is $\tilde{Y} = \azul{Y_1} \!\cup \!\azul{Y_2}$. However, for control design purposes, we consider the output admissible set to be $Y\! =\! \{y: \|y\|_{\infty} \leq 2\}$ with an associate avoidance cost functional to avoid the non-feasible output region $O \!=\! Y\! -\! \tilde{Y}$. The system is constrained to $\tilde{Y}$ if the disjoint constraint $y \!\in\! \azul{Y_1}\! \vee\! y\! \in\! \azul{Y_2}$ holds. \azul{Notice that in this example it is easier to define the constraint $y\! \in\! \tilde{Y}$ instead of $y\! \notin\! O$. For that, we need a penalty function $F(y, \tilde{Y})$ that is greater than zero when $y\! \notin \!\tilde{Y}$, and zero otherwise. Considering the disjoint constraint $y\! \in \!\azul{Y_1}\! \vee y\! \in\! \azul{Y_2}$ to obtain $y\! \in\! \tilde{Y}$, such a function can be obtained from the product of the functions describing $Y_1$ and $Y_2$ \citep{Hermans2018},} \mage{yeilding $F(y, \azul{\tilde{Y}}) = (\mathrm{max}\{0,g_1(y)\}\mathrm{max}\{0,g_2(y)\})^2$,} where $g_1(y)\! =\! (y-y_{c_1})'E_1(y - y_{c_1})\! -\! 1\! +\! \gamma_1$ and $g_2(y)\! = \!(y-y_{c_2})'E_2(y - y_{c_2})\! -\! 1\! +\! \gamma_2$, with $\gamma_1\! =\! \gamma_2\! =\! 0.15$ being a constant to define a safety region around $O$.

The linear model required for prediction is obtained from \eqref{eq:ballandplatemodel} through the linearization around the equilibrium condition $(x_{eq}, u_{eq})$ with $x_{eq} = O_{8,1}$ and $u_{eq} = O_{2,1}$. Afterward, the linearized system is discretized considering the Euler approximation with sampling time $T_s = 0.25$ s. The system starts in the initial condition $x(0) = [0~ -1.2 ~ 0~ 0~ 0~ 0~ 0~ 0]'$, and it is required to reach \azul{two distinct target outputs, $y_{t_1} = [-0.85 ~ 0.60]'$ and $y_{t_2} = [1 ~ -0.25]'$}. The horizon prediction is $N = 8$, the weighting matrices are $Q = \mathrm{diag}\{2,2,1,1,10,10,50,50\}$ and \azul{$R = \mathrm{diag}\{0.01, 0.01\}$}, the stabilizing control gain $K$ is the \ver{linear quadratic regulator} gain, and the terminal cost $P$ is obtained solving the associated Riccati equation. Furthermore, for the avoidance cost, we consider $\mu = 10^5$\mage{,} while the offset cost is defined as $V_{of}(y_a, y_t) = \|y_a - y_t\|^2_{\kappa}$ with $\kappa = \mathrm{diag}\{10^4, 10^4\}$.

\begin{figure}[!htbp]	
	\centering{
		\small
		\def\svgwidth{0.3\textwidth}
		{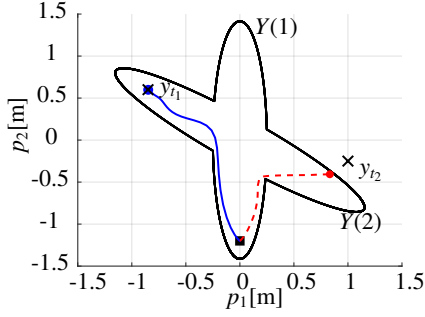}
		\normalsize
  \caption{\azul{Complete trajectory performed by the ball over the non-convex plate composed by the union of two ellipses when tracking the targets $y_{t_1}$ and $y_{t_2}$, which are depicted, respectively, by the blue and the red-dashed lines. In the figure, the black cross-markers denote the desired set-points.}}		
		\label{fig:ballplate2D}}
\end{figure}

\begin{figure*}[!t]	
		\centering{
		\footnotesize
		\def\svgwidth{0.85\linewidth}
		{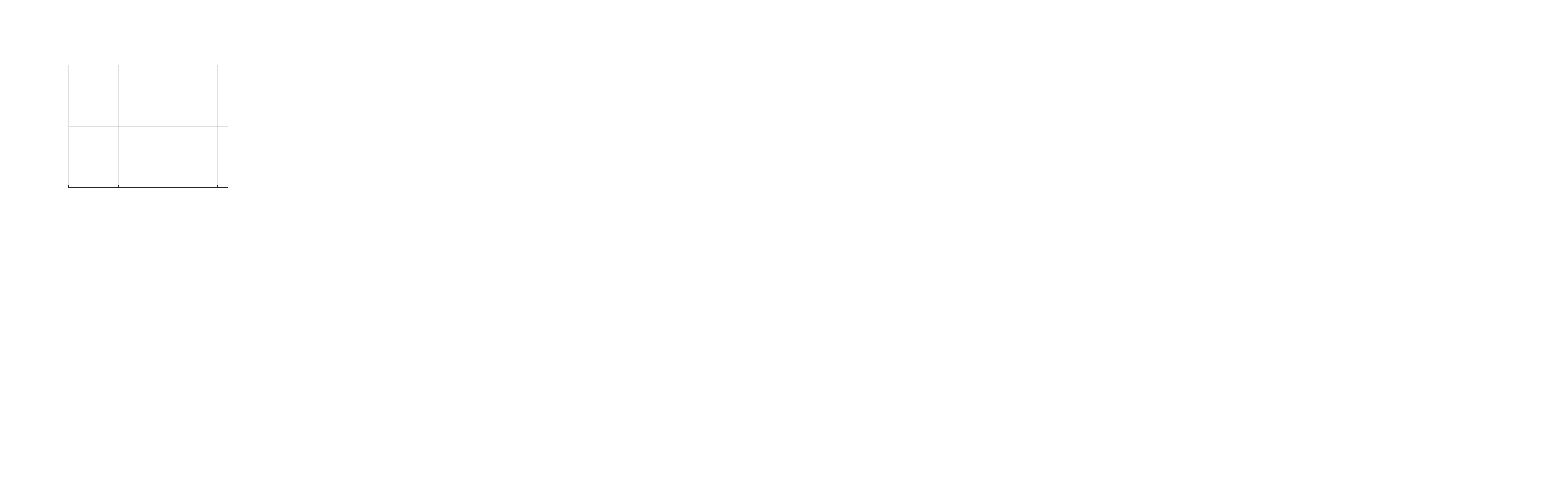}
		\normalsize
		\caption{\mage{(a) Absolute output tracking error with $e_{p_1} = p_1 - p_{1t}$ and $e_{p_2} = p_2 - p_{2t}$ \azul{for the targets $y_{t_1}$ and $y_{t_2}$}. In the figure, it is illustrated the relation between error reduction and avoidance. (b) Time evolution of the ball position ($p_1$ and $p_2$) and the plate angles ($\theta_1$ and $\theta_2$) during the tracking of the \azul{provided targets}. In the figures, the dashed lines denote the artificial references.  (c) Time evolution of the desired angular accelerations ($a_1$ and $a_2$), the manipulated variables of the system.}}
		\label{fig:ballplatejointplot}}
\end{figure*}

Figure \ref{fig:ballplate2D} shows the shape of the non-convex plate defined by the \ver{union} of the ellipses $\azul{Y_1}$ and $\azul{Y_2}$. In addition, it depicts the output trajectory performed by the closed-loop system from the initial condition \azul{towards both desired outputs $y_{t_1}$ and $y_{t_2}$} (indicated by the black x-shaped \azul{markers}). \azul{For the target $y_{t_1}$,} notice that the position of the ball starts following the ellipse $\azul{Y_1}$ contour with a safe distance until there is a clear path towards the output target. This behavior can be seen in more detail in \azul{the left plots of} Figure \ref{fig:ballplatejointplot}(a)\mage{. Notice that,} especially on the \azul{upper-left} graphic, the tracking error decay rate reduces up to the point where it becomes almost constant. After the cornering point where the ellipses intersect each other, the decay rate increases again. Unlike any controller designed only for set-point tracking, the avoidance term disturbs the asymptotic convergence when necessary. However, since the closed-loop system is ISS with respect to the avoidance cost, at some point the effect of the avoidance may vanish and the asymptotic convergence to an admissible steady-output point is reached again. \azul{Analyzing the trajectory to the target $y_{t_2}$, it is possible to observe the second condition of Theorem \ref{theoremAntonio} since the target is unfeasible, i.e., $y_{t_2} \notin Y_r \cap \tilde{Y}$. As expected, the system converges to an admissible output that minimizes $V_{of}(y_a, y_{t_2}) + V_{av}(\bm{y}, y_a, O)$. This behavior is corroborated in the right plots of Figure \ref{fig:ballplatejointplot}(a) by noticing that the tracking error converges to a positive value instead of zero.}

Furthermore, Figure \ref{fig:ballplatejointplot}(b) shows \azul{for both targets} the evolution of the ball position and the plate orientation \azul{together with the artificial references denoted by the dashed lines}. Besides showing the convergence to the desired position, this figure also depicts the angles stably converging to the steady condition of null orientation of the plate. Finally, Figure \ref{fig:ballplatejointplot}(c) depicts the calculated control inputs \azul{for both targets}. It can be seen in the figure that the controller was able to perform set-point tracking while satisfying the constraints imposed on the control action.

\begin{remark}
	\ver{On one hand, when compared to the results obtained in \citet{Cotorruelo2021}, the proposed controller is less computationally efficient since its control law is derived from the solution of a nonlinear programming problem while in \citet{Cotorruelo2021} it only requires the solution of a second order cone programming problem. On the other hand, unlike \citet{Cotorruelo2021}, the results obtained in this work still hold if a convexifying homeomorphism of the output space does not exist. These results are expected since the \mage{proposed} framework was developed to provide avoidance features.}
\end{remark}

\subsection{UAV navigation in cluttered environment}

\mage{Consider the quadrotor UAV described in \citep{Raffo2011}, where the position of the body frame's origin expressed in the inertial frame is given by $\xi = [x^{\mathcal{I}}  y^{\mathcal{I}} z^{\mathcal{I}}]'$ and its attitude by $\eta = \left[\phi \; \theta \; \psi\right]'$.} Thus, the generalized coordinates describing the quadrotor UAV motion can be chosen as $q = \left[\xi' \; \eta' \right]'$, which leads to the state-vector $x = \left[q' \; \dot{q}' \right]'$. Moreover, the system inputs are $u = \left[f_1 \; f_2 \; f_3 \; f_4 \right]'$, with $f_i$ being the thrust force generated by the $i$-th rotor, and the output-vector is $y = [x^{\mathcal{I}}  y^{\mathcal{I}} z^{\mathcal{I}} \psi]'$. Aiming to better emulate the vehicle dynamics during the simulation, a more complete dynamic model is considered taking into account the coupling between translational and rotational dynamics due to the displacement 
between the quadrotor's geometric center and its center of mass\mage{.} However, for control design purposes, this displacement is neglected, \mage{and, for prediction, it is used a model linearized around the equilibrium} $q_{eq} = [{\xi_{eq}}' ~ {\eta_{eq}}']'$ and $u_{eq} = \left[f_{1,eq} ~ f_{2,eq} ~ f_{3,eq} ~ f_{4,eq}\right]'$. For the sake of simplicity, the models are omitted. 

Since the \mage{proposed} control strategy requires a discrete dynamical model, a sampling time of $T_s = 0.01$ \mage{s} is considered for discretization using the Euler approximation. \mage{Also}, to access the capacity of avoiding non-feasible regions, we consider a $48 \times 30 \times 20$ \mage{m} map with \azul{7} rectangle-shaped obstacles obstructing the UAV workspace. A 3D lidar-like sensor is emulated to detect the obstacles within a spherical range of \mage{$4$ m} based on the quadrotor UAV global position and the environment map. However, it is \mage{noteworthy} that, from the control algorithm standpoint, the obstacles are \ver{previously unknown} and perceived only by the obstacle detection system. Hence, they are only avoided when inside the sensor range. Based on the sensor information, the \ver{obstacles} are defined in execution time as a \mage{$2$ m} radius sphere centered in \ver{the closest} point measured in the boundary of the obstacle. Therefore, considering \mage{$g(y,O_i) = - (y-y_{c_i})'I_3(y - y_{c_i}) + 2^2$}, the penalty function can be defined as \azul{$F(y,\azul{O_i}) = \mathrm{max}\{0,g_j(y, \azul{O_i})\}^2$}.

\mage{Here}, we consider the terminal equality constraint version of the controller (see Remark \ref{remark1}) with horizon $N = 50$ and the weighting matrices $Q = \mathrm{diag}\{1, 1, 1, 0.1, 0.1,$ $1, 1,1,1,10,10,1\}$ and $R = \mathrm{diag}\{10, 10, 10, 10\}$. As for the offset and avoidance costs, $V_{of}(y_a, y_t) = \|y_a - y_t\|^2_{\kappa}$ with $\kappa = \mathrm{diag}\{4000, 4000, 30000, 4000\}$ and $\azul{\mu_i} = 40000$\azul{, for all $O_i$}. \mage{Also}, the admissible input set is \mage{$U = \{(f_1,f_2,f_3,f_4) \in \mathbb{R}^4 : 0 \leq f_i \leq 12, \; i=1,\ldots, 4\}$}, and the admissible state set is defined accordingly to the map dimensions and the system operational conditions $X = \{[\xi'~\eta'~\dot{\xi'}~\dot{\eta}']' : \azul{[-24 ~ -15 ~ 0]'} \leq \xi \leq [24 ~ 15 ~ 20]', |\eta| \leq [\pi/2 ~ \pi/2 ~ \pi]', |\dot{\xi}| \leq [5 ~ 5 ~ 5]', |\dot{\eta}| \leq [\pi/2 ~ \pi/2 ~ \pi/2]' \}$.
\begin{figure}[!hbp]	
	\centering{
		\includegraphics[width=0.35\textwidth]{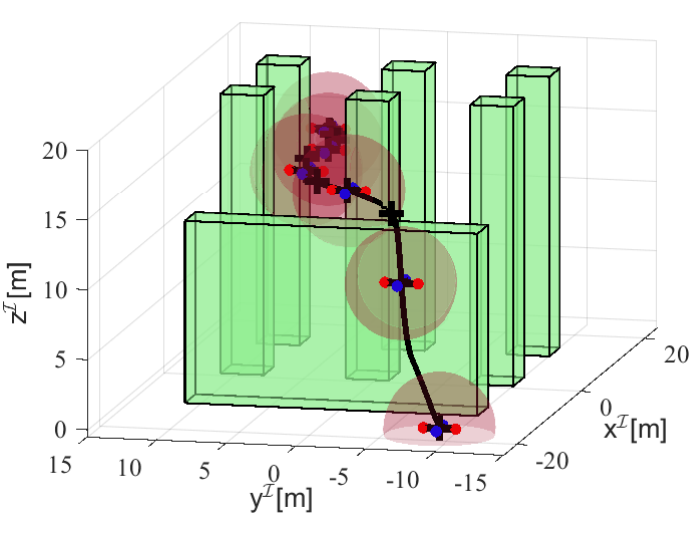}
		\caption{\ver{Trajectory performed by the UAV (black line) to safely complete a given task while avoiding \azul{7} previously unknown obstacles, which are denoted by the green rectangle-shaped objects. In the figure, the UAV is denoted as a black cross with blue and red spheres in its extremities and the light red sphere around the UAV denotes the range in which obstacles can be detected.}}}		
		\label{fig:quad3Dplot}
\end{figure}
\begin{figure}[!b]	
	\centering{
		\small
		\def\svgwidth{0.27\textwidth}
		{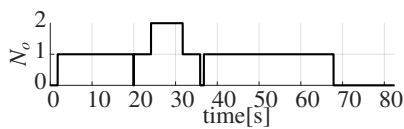}
		\normalsize
		\caption{\azul{Number of obstacles being perceived by the 3D lidar-like sensor emulated to detect obstacles within a given range.}}
		\label{fig:quadobstacles}}
\end{figure}
\begin{figure*}[!htbp]	
	\centering{
		\footnotesize
		\def\svgwidth{0.85\linewidth}
		{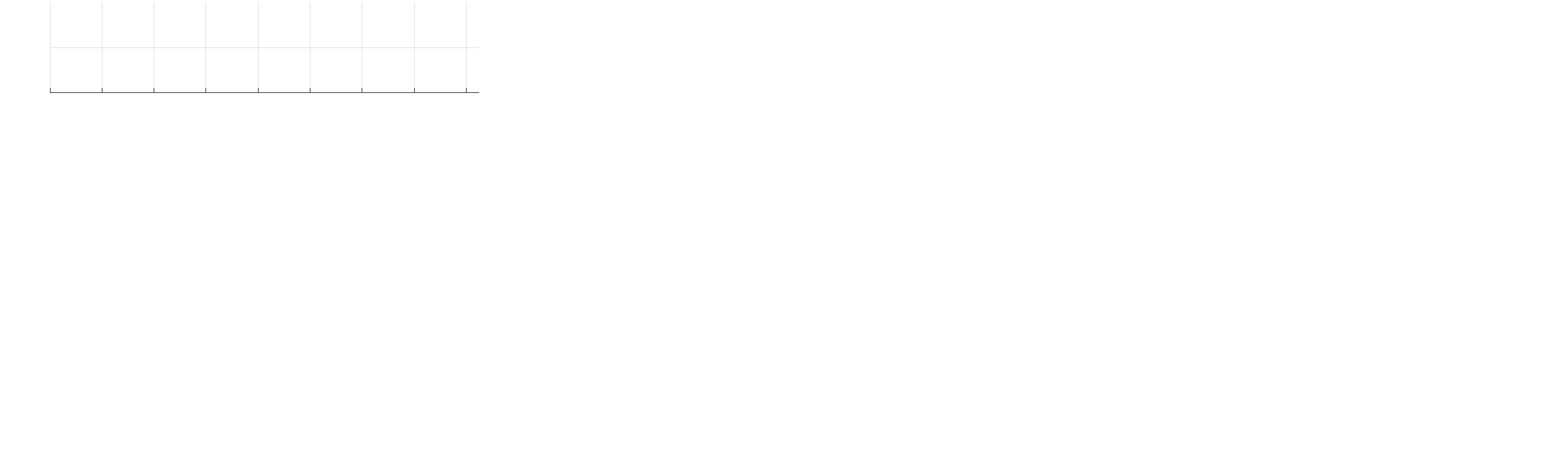}
		\normalsize
		\caption{\mage{(a) Absolute output tracking error illustrating the relation between error reduction and avoidance. In the figure, $e_x = x - x_t$, $e_y = y - y_t$, and $e_z = z - z_t$. (b) Time evolution of the quadrotor UAV orientation ($\phi$, $\theta$, and $\psi$) during the execution of the provided task. In the figures, the dashed lines denote the artificial references.  (c) Time evolution of the applied lift forces to the quadrotor UAV, with $f_i$ being the lift force of the $i$-th propeller.}}
		\label{fig:quadjointplot}}
\end{figure*}

In this simulation scenario, the quadrotor UAV is required to go autonomously from its initial position $[-17 ~ -12 ~ 0]'$ to the desired position $[18.75 ~ 8 ~ 15]'$, depicted in the upper part of Figure \ref{fig:quad3Dplot}, with $\psi = 0$. It is expected that the proposed set-point tracking MPC with avoidance features will find an alternative path around the \ver{obstacles} by means of the artificial variables. In fact, Figure \ref{fig:quad3Dplot} shows the vehicle after it reaches the goal and the alternative path performed to reach the desired target. Then, the use of the proposed control strategy provides the system with two interesting features. First, in the presence of \ver{obstacles}, the controller autonomously finds a path around them by means of the artificial variables. Second, it provides feasible intermediary equilibrium points when the required target is not reachable in $N$ steps by the system due to its dynamics and constraints. \ver{It is worthwhile mentioning that the obstacles shape and spatial distribution has direct impact on the ability to obtain a global solution to the optimal control problem. However, the proposed formulation does not limit neither the shape nor the spatial distribution of the regions to be avoided. Furthermore, by perceiving the obstacles within a given range as spheres centered in the closest points measured in the boundaries of each obstacle, from the control algorithm standpoint, the number and position of the perceived obstacles change over-time despite the environment being static.} \azul{Figure \ref{fig:quadobstacles} exemplifies this behavior by showing the number of detected obstacles at each time instant.}

The avoidance process can also be verified by looking at the absolute position error. In fact, the control algorithm gives up performance to achieve safe navigation, which can be seen in Figure \ref{fig:quadjointplot}(a) where the error stops decreasing at some points even before converging to zero. Moreover, Figure \ref{fig:quadjointplot}(b) presents the time evolution of the quadrotor UAV orientation stably converging to an equilibrium value. This is expected since at the end of the task execution the UAV will be in hovering flight mode due to the equilibrium condition of the artificial variables. Notice that, since the artificial steady-output is defined in the output level, there is no need to require a reference for $\phi$ and $\theta$. In fact, the artificial orientation\azul{, which are depicted in Figure \ref{fig:quadjointplot}(b) as dashed lines,} is obtained from the artificial position by means of the model constraints. Finally, Figure \ref{fig:quadjointplot}(c) shows the control signals applied to the quadrotor UAV.

\vspace{-2mm}
\section{Conclusions}

In this work, it was proposed a linear model predictive control strategy able to perform set-point tracking while avoiding non-feasible regions inside the prior known admissible space. For that, three main ingredients were considered: i) artificial variables to represent artificial steady conditions; ii) offset cost functional playing the role of a steady-state target optimizer; and iii) an avoidance cost functional avoiding the system evolution to lie inside non-feasible regions. It was shown that, under mild conditions, the closed-loop system is ISS with respect to the avoidance cost. \ver{Thus, it was possible to demonstrate that the closed-loop system is stable and recursively feasible.} Two numerical examples were proposed to analyze the behavior of the proposed control strategy. First, a ball-on-plate system with a non-convex plate is considered to show how a problem with non-convex admissible sets could be handled within the proposed formulation. Second, it is considered the autonomous navigation problem of a quadrotor UAV in a cluttered environment with \ver{previously unknown} obstacles. \ver{In future works, the proposed control strategy will be extended to the nonlinear case, robustification strategies will be addressed to handle uncertainties, and computationally efficient solutions will be proposed aiming at real-world applications.}

\section*{Appendix. Proof of Lemmas 1 to 7}

\begin{proof}(\textbf{Lemma 1})
	\ver{Consider that $(x_a^O, u_a^O, y_a^O)$ is the optimal solution to $P_N^O(x, y_t, \azul{O_i})$. Then \mage{$V_N^O(x, y_t, \azul{O_i}) = V_{of}(y_a^O, y_t) + V_{av}(\bm{y}, y_a^O, \azul{O_i})$.} This Lemma will be proved by contradiction, extending the results of \citet[Lemma 1]{Limon2018} for the case with avoidance. For that, assume now that the stationary point is not optimal, i.e., $(x_a^O,u_a^O) \neq (x_s, u_s)$. Let us define \mage{$(\tilde{x}_a, \tilde{u}_a) = \beta(x_a^O,u_a^O) + (1-\beta)(x_s, u_s)$} with $\beta \in [0,1]$. Since both $(x_s, u_s)$ and $(x_a^O,u_a^O)$ are in $Z_s$, and this set is convex, then a convex combination of these points, $(\tilde{x}_a, \tilde{u}_a)$, is also in $Z_s$.}
	
	\ver{Considering Assumptions \ref{assump5d} and \ref{assump6}, it is possible to obtain a convex cost functional that superiorly bounds the non-convex cost $V_{of}(y_a, y_t) + V_{av}(\bm{y}, y_a, \azul{O_i})$. Then, we can define \mage{$V_B(y_a, y_t) = V_{of}(y_a^O, y_t) + \azul{S},$ for any bound $\azul{S}$, such that}}
    \begin{equation}
        \label{eq:lemma1den}
	\ver{V_B(\tilde{y}_a,y_t) \leq V_B(y_a^O,y_t)}
    \end{equation} 
    \ver{for every $\beta$. In other words, since the system is not at the optimal point $(x_s, u_s)$, it is more convenient to move towards $(\tilde{x}_a, \tilde{u}_a)$ than to remain in $(x_a^O,u_a^O)$. }
	
	\ver{Let $\bm{\tilde{u}}$ be a feasible control sequence that drives the system from $(x_a^O,u_a^O)$ to $(\tilde{x}_a, \tilde{u}_a)$. This sequence is such that, the j-th element is given by $\tilde{u}(j) = K(\tilde{x}(j) - \tilde{x}_a) + \tilde{u}_a$ and $\tilde{x}(j+1) = A\tilde{x}(j) + B\tilde{u}(j)$, with $\tilde{x}(0) = x_a^O$. Then, the cost to drive the system to $(\tilde{x}_a, \tilde{u}_a)$ in $N$ steps is} 
	\begin{align*}
		\ver{V_N(x_a^O, y_t, \azul{O_i}) =} &\ver{\sum\limits_{j=0}^{N-1}\|\tilde{x}(j) -  \tilde{x}_a\|_{Q}^2 + \| K(\tilde{x}(j) - \tilde{x}_a) \|^2_{R}}  \\
		&~  \ver{+ \|\tilde{x}(N) -  \tilde{x}_a\|_{P}^2 +  V_{of}(\tilde{y}_a, y_t) + V_{av}(\bm{\tilde{y}}, \tilde{y}_a, \azul{O_i})} \\
		\ver{\leq} & \ver{~ \|x_ a^O -  \tilde{x}_a\|_{P}^2 +  V_{of}(\tilde{y}_a, y_t) + \azul{S}} \\
		\ver{\leq} & \ver{~ (1-\beta)^2\|x_ a^O -  x_s\|_{P}^2 +  V_{of}(\tilde{y}_a, y_t) + \azul{S}.}
	\end{align*}
	
	\ver{Now define $W(\beta) = (1-\beta)^2\|x_ a^O -  x_s\|_{P}^2 +  V_{B}(\tilde{y}_a, y_t)$ and notice that for $\beta = 1$, $W(1) = V_{B}(y_a^O, y_t)$. Taking the partial derivative with respect to $\beta$ and evaluating it for $\beta = 1$, we obtain \mage{${\left.\partial W/\partial \beta\right|_{\beta=1}=g^{O^{\prime}}(y_a^O, y_t)}$,} with $g^{O^{\prime}}(y_a^O, y_t) \in \partial V_{B}(y_a^O, y_t)$, where $\partial V_{B}(y_a^O, y_t)$ is defined as the subdifferential of $V_{B}(y_a^O, y_t)$.} 
	
	\ver{From convexity and from \eqref{eq:lemma1den}, \mage{$\left.\partial W/\partial \beta\right|_{\beta=1}=g^{O^{\prime}}(y_a^O, y_t) \geq V_{B}(y_a^O, y_t) - V_{B}(\tilde{y}_a, y_t) > 0$.} This means that there exists a value of $\beta \in [0,1)$ such that $V_B(\tilde{y}_a, y_t)$ is smaller than the value of the cost $V_B(\tilde{y}_a, y_t)$ for $\beta = 1$, which is $V_{B}(y_a^O, y_t)$. This contradicts the optimality of the solution of $P_N^O(x, y_t,\azul{O_i})$. Then, it has to be $(x_a^O,u_a^O) = (x_s, u_s)$, with $(x_s, u_s)$ being the minimizer of $V_{of}(y_a, y_t) + V_{av}(\bm{y}, y_a, \azul{O_i})$, which concludes the proof.}
\end{proof}

\begin{proof}(\textbf{Lemma 2})
	\ver{This lemma extends the results of \citet[Lemma 4]{DJorge2020} for the case with avoidance. Thus, following a similar analysis, due to convexity, $e(x)$ is a continuous function \citep[Theorem A.23]{Rawlings2009}. Moreover, let us consider these two cases.} 
	\begin{enumerate}
		\item \ver{$\|e(x)\| = 0$ \mage{iff} $x = x_s$. In fact, i) if $e(x) = 0$, then $x = x_a^O$, and from Lemma \ref{lemma:Limon2018}, this implies that $x_a^O = x_s$; ii) if $x = x_s$, then by optimality $x_a^O = x_s$, and then $x = x_a^O$. Then, $\|e(x)\| = 0$.}
		\item \ver{$\|e(x)\|>0$ for all $\|x-x_s\|>0$. In fact, for any $x\neq x_s$, $\|e(x)\|\neq 0$ and moreover $\|x-x_s\|>0$. Then, $\|e(x)\|>0$.}
	\end{enumerate}
	\ver{Therefore, since $X_N(\Omega_t)$ is compact \citep[Chapter 5 - Lemma 6]{Vidyasagar1993}, there exists a $\mathcal{K}$-function $\alpha_e(\cdot)$ such that $\|e(x)\| \geq \alpha_e(\|x-x_s\|)$ on $X_N(\Omega_t)$, which concludes the proof.}
\end{proof}

\begin{proof}(\textbf{Lemma 3})
	For a feasible state $x \in X_N(\Omega_t)$ at time $k$, the optimal cost functional is $V_N^O(x, y_t, \azul{O_i})$, with the decision variables $(\bm{u}^O,  x_{a}^O, u_{a}^O)$ being the optimal solution of the problem $P_N^O(x, y_t, \azul{O_i})$. The obtained optimal control sequence $\bm{u}^O = (u^O(0), u^O(1), \cdots, u^O(N-1))$ is associated with the optimal predicted state sequence $\bm{x}^O = (x^O(0), x^O(1), \cdots, x^O(N-1), x^O(N))$ with $x^O(N) \in \Omega_t$.
	Defining an auxiliary feasible input sequence $\tilde{\bm{u}} = (u^O(1), \cdots, u^O(N-1), K(x^O(N) - x_a^O) + u_a^O)$, an auxiliary feasible artificial state $\tilde{x}_a = x_a^O$, and an auxiliary feasible artificial input $\tilde{u}_a = u_a^O$, the state sequence associated to $(\tilde{\bm{u}}, \tilde{x}_a, \tilde{u}_a)$ starting from $x(k+1) = Ax(k) + Bu^O(0)$ is given by \mage{$\tilde{\bm{x}} = (x^O(1), \cdots, x^O(N), x(N+1))$,} with $x(N+1) = A x^O(N) + B(K(x^O(N) - x_a^O) + u_a^O)$. Since $(x^O(N), x_a^O, u_a^O) \in \Omega_t^a$, the control action $K(x^O(N) - x_a^O) + u_a^O$ is admissible and the terminal state $x(N+1)$ is feasible due to the positive invariance of $\Omega_t^a$, i.e., $(x(N+1), \tilde{x}_a, \tilde{u}_a) \in \Omega_t^a$. Therefore, $x(k+1) \in X_N(\Omega_t)$, proving that the closed-loop system is recursively feasible.
\end{proof}

\begin{proof}(\textbf{Lemma 4})
	\mage{From the shifted value function definition, the suboptimality of the feasible law $\bm{u}_f$, i.e., $u_f(j) \in U ~\forall j \geq 0$, and the boundedness of the avoidance function implying $V_{av}(\bm{y}, y_a, \azul{O_i}) - \azul{S} \leq 0$, it holds that}
	\begin{align}
		&~ \sum\limits_{j=0}^{N-1}\|x^O(j) -  x_a^O\|_{Q}^2 + \| u^O(j) - u_a^O \|^2_{R} + \|x^O(N) -  x_a^O\|_{P}^2 \nonumber \\
		&~ +  V_{of}(y_a^O, y_t) + V_{av}(\bm{y}^O, y_a^O, \azul{O_i}) - \azul{S} \nonumber
        \end{align}
        \begin{align}
		\leq & \sum\limits_{j=0}^{N-1}\|x(j) -  x_a\|_{Q}^2 + \| u_f(j) - u_a \|^2_{R} + \|x(N) -  x_a\|_{P}^2 +  V_{of}(y_a, y_t) \nonumber  \\
		= & ~J(x).
	\end{align}
        Since $J(x)$ is a locally bounded continuous function with $J(x_s) = 0$ (Lemma \ref{lemma:Limon2018}), then there exists a $\mathcal{K}_{\infty}$-function $\alpha_c(\cdot)$ such that $J(x) \leq \alpha_c(\|x - x_s\|)$, for all $x \in X_N(\Omega_t)$  \citep{Rawlings2009}.
\end{proof}

\begin{proof}(\textbf{Lemma 5})
	From Assumptions \ref{assump5} to \ref{assump5b}, there is a $\mathcal{K}_{\infty}$-function $\alpha(\|x - x_a\|)$ such that \mage{$\|x -  x_a\|_{Q}^2 + \| u - u_a \|^2_{R} +  \|x -  x_a\|_{P}^2 \geq \alpha(\|x - x_a\|).$} Following the definition of the \mage{$V_s(x, y_t, \azul{O_i})$}, we have	
	\begin{align}
		&\sum\limits_{j=0}^{N-1}\|x(j) -  x_a\|_{Q}^2 + \| u(j) - u_a \|^2_{R} + \|x(N) -  x_a\|_{P}^2  \\
		&~ +  V_{of}(y_a, y_t) + V_{av}(\bm{y}, y_a,\azul{O_i}) - \azul{S}  \mage{\geq \sum\limits_{i=0}^{N-1} \alpha(\|x - x_a\|) - \azul{S}.} \nonumber
	\end{align}
	
	Finally, based on Lemma \ref{lemma:error} and on the optimality principle,
	\begin{equation}
		\sum\limits_{i=0}^{N-1} \alpha(\|x - x_a\|) - \azul{S} \geq \hat{\alpha}_b(\|x - x_a^O\|) - \azul{S} \geq \alpha_b(\|x - x_s\|) - \azul{S},
	\end{equation}
	with $\alpha_b(r) = \hat{\alpha}_b \circ \alpha_e(r)$.
\end{proof}

\begin{proof}(\textbf{Lemma 6})
	Let $\bm{u}^O = (u^O(0), u^O(1), \cdots, u^O(N-1))$ be an optimal control sequence, $\tilde{\bm{u}} = (u^O(1), \cdots, u^O(N-1),$ $K(x^O(N) - x_a^O) + u_a^O)$ be an auxiliary feasible input sequence, $\bm{y}^O = (y^O(0), y^O(1), \cdots, y^O(N))$ be an optimal output sequence, $\tilde{\bm{y}} = (y^O(1), \cdots, y^O(N), y(N+1))$ be an auxiliary feasible output sequence, $\tilde{x}_a = x_a^O$ be an auxiliary feasible artificial state, $\tilde{y}_a = y_a^O$ be an auxiliary feasible artificial output, and $\tilde{u}_a = u_a^O$ be an auxiliary feasible artificial input \ver{with the triplet $(\tilde{x}_a, \tilde{u}_a, \tilde{y}_a)$ being the feasible solution to the one-step ahead optimization problem.} \mage{Further, let $x(k+1) = Ax(k) + Bu^O(0)$.} 
	
	Comparing, at $k+1$, $V_s(x(k+1), y_t, \azul{O_i};\tilde{\bm{u}}, \tilde{x}_a, \tilde{u}_a)$ and $V_s^O(x(k), y_t, \azul{O_i})$, it is possible to obtain	
	\begin{align}
		V_s(&x(k+1), y_t, \azul{O_i};~\tilde{\bm{u}}, \tilde{x}_a, \tilde{u}_a) - V_s^O(x(k), y_t, \azul{O_i}) =  \nonumber \\
		& -\|x^O-x_a^O\|_Q^2 -\|u^O(0)-u_a^O\|_R^2 - \|x^O(N) - x_a^O\|_P^2\nonumber \\
		& - V_{of}(y_a^O, y_t) - V_{av}(\bm{y}^O, y_a^O, \azul{O_i}) + \azul{S} + \|x^O(N) - \tilde{x}_a\|_Q^2\nonumber \\
         & + \|K(x^O(N) - \tilde{x}_a)\|_R^2 + \|x(N+1) - \tilde{x}_a\|_P^2 + V_{of}(\tilde{y}_a, y_t) \nonumber \\
         & + V_{av}(\tilde{\bm{y}}, \tilde{y}_a, \azul{O_i}) - \azul{S} \nonumber \\
        \leq & \ver{-\|x-x_a^O\|_Q^2 -\|u^O(0)-u_a^O\|_R^2} \nonumber \\
        & + V_{av}(\tilde{\bm{y}},\tilde{y}_a, \azul{O_i}) - V_{av}(\bm{y}^O,y_a^O, \azul{O_i}).
	\end{align}
	Based on the boundedness of the avoidance function, $V_{av}(\tilde{\bm{y}},\tilde{y}_a, \azul{O_i}) - V_{av}(\bm{y}^O,y_a^O, \azul{O_i}) \leq \azul{S}$, on the optimality principle, $V_s^O(x(k+1), y_t, \azul{O_i}) \leq V_s(x(k+1), y_t, \azul{O_i};\tilde{\bm{u}}, \tilde{x}_a, \tilde{u}_a)$, and on Lemma \ref{lemma:error}, there is a $\mathcal{K}_{\infty}$-function $\alpha(\|x - x_s\|)$ such that
	\begin{align}	
		V_s^O(x(k+1), y_t, \azul{O_i}) - & V_s^O(x(k), y_t, \azul{O_i}) \leq -\hat{\alpha}(\|x - x_a^O\|) + \azul{S} \nonumber \\
		& \leq  - \alpha(\|x - x_s\|) + \azul{S}, 
	\end{align}
	with $\alpha(r) = \hat{\alpha} \circ \alpha_e(r)$.
\end{proof}

\begin{proof}(\textbf{Lemma 7})
	
	This proof can be obtained following the steps considered in \citet{Jiang2001} to proof Lemma 3.5. Thus, rewriting Lemma \ref{lemma4} and considering Lemma \ref{lemma2}, we have
	\begin{align}
		V_s^O(x(k+1), y_t, \azul{O_i}) - & V_s^O(x(k), y_t, \azul{O_i}) \label{formulation1:eqbeta} \\
		& \leq -\alpha\circ\alpha_c^{-1}(V_s^O(x(k), y_t, \azul{O_i})) + \azul{S}. \nonumber 
	\end{align}
	Without loss of generality, for $\hat{\alpha} = \alpha \circ \alpha_c^{-1}$, we assume \ver{$(id - \hat{\alpha})(\cdot)$} to be a $\mathcal{K}$-function. Let $\rho$ be any $\mathcal{K}_{\infty}$-function such that \ver{$(id - \rho)(\cdot)$} is a $\mathcal{K}_{\infty}$-function and consider the set defined by \mage{$\mathcal{D} = \{x: V_s^O(x(k), y_t, \azul{O_i}) \leq b\}$,} where $ b = \hat{\alpha}^{-1}\circ\rho^{-1}(\azul{S})$.
	
	\begin{claim}
		If there is some $k_0 \in \mathbb{Z}_{>0}$ such that $x(k_0)\in \mathcal{D}$, then $x(k) \in \mathcal{D}$ for all $k\geq k_0$.
	\end{claim}
	\begin{proof}
		Assume $x(k_0) \in \mathcal{D}$. Then $V_s^O(x(k_0), y_t, \azul{O_i})\leq b$, i.e., $\rho \circ \hat{\alpha}(V_s^O(x(k_0), y_t, \azul{O_i}) \leq \azul{S}$. By \eqref{formulation1:eqbeta}, 
		\begin{equation}
			V_s^O(x(k_0+1), y_t, \azul{O_i}) \leq (id - \hat{\alpha})(V_s^O(x(k_0), y_t, \azul{O_i})) + \azul{S},
		\end{equation}
		and since $id - \hat{\alpha}$ is a $\mathcal{K}$-function, we have
		\begin{align}
			V_s^O(x(k_0+1), y_t, \azul{O_i}) & \leq(id-\hat{\alpha})(b) + \azul{S} \nonumber\\
			&=-(id-\rho)\circ\hat{\alpha}(b) + b \leq b.
		\end{align}
		By induction, it is possible to show that $V_s^O(x(k_0 + j), y_t, \azul{O_i})\leq b$ for all $j\in \mathbb{Z}_{>0}$, that is, $x(k)\in \mathcal{D}$ for all $k\geq k_0$.
	\end{proof}
	
	We now let $j_0 = \mathrm{min}\{k\in\mathbb{Z}_{>0}:x(k) \in \mathcal{D}\} < \infty$. Then, it follows from the above conclusion that $V_s^O(x(k), y_t, \azul{O_i})\leq \hat{\gamma}(\azul{S})$ for all $k\geq j_0$, where $\hat{\gamma}(r) = \hat{\alpha}^{-1}\circ\rho^{-1}(r)$. For $k < j_0$, it holds that $\rho \circ \hat{\alpha}(V_s^O(x(k), y_t, \azul{O_i})) > \azul{S}$, and hence
	\begin{align}
		V_s^O&(x(k+1), y_t, \azul{O_i}) - V_s^O(x(k), y_t, \azul{O_i})) \\
		& \mage{\leq -\hat{\alpha}(V_s^O(x(k), y_t, \azul{O_i})) + \azul{S}\leq -(id - \rho)\circ\hat{\alpha}(V_s^O(x(k), y_t, \azul{O_i}))} \nonumber
	\end{align}
	By a comparison lemma \citep{Jiang2002}, there exists some $\mathcal{KL}$-function $\hat{\beta}$ such that \mage{$V_s^O(x(k), y_t, \azul{O_i}) \leq \hat{\beta}(V_s^O(x(0), y_t, \azul{O_i}),k)$} for all $0\leq k < j_0$. Thus, 
	\begin{equation}
		V_s^O(x(k), y_t, \azul{O_i}) \leq \mathrm{max}\{\hat{\beta}(V_s^O(x(0), y_t, \azul{O_i}),k),\hat{\gamma}(\azul{S})\},    
	\end{equation}
        for all $k \in \mathbb{Z}_{>0}$.
\end{proof}

\bibliography{mybibfile}

\end{document}

%% file: images/ballandplate_final/2Dplot.pdf_tex
\begingroup%
  \makeatletter%
  \providecommand\color[2][]{%
    \errmessage{(Inkscape) Color is used for the text in Inkscape, but the package 'color.sty' is not loaded}%
    \renewcommand\color[2][]{}%
  }%
  \providecommand\transparent[1]{%
    \errmessage{(Inkscape) Transparency is used (non-zero) for the text in Inkscape, but the package 'transparent.sty' is not loaded}%
    \renewcommand\transparent[1]{}%
  }%
  \providecommand\rotatebox[2]{#2}%
  \newcommand*\fsize{\dimexpr\f@size pt\relax}%
  \newcommand*\lineheight[1]{\fontsize{\fsize}{#1\fsize}\selectfont}%
  \ifx\svgwidth\undefined%
    \setlength{\unitlength}{420bp}%
    \ifx\svgscale\undefined%
      \relax%
    \else%
      \setlength{\unitlength}{\unitlength * \real{\svgscale}}%
    \fi%
  \else%
    \setlength{\unitlength}{\svgwidth}%
  \fi%
  \global\let\svgwidth\undefined%
  \global\let\svgscale\undefined%
  \makeatother%
  \begin{picture}(1,0.75)%
    \lineheight{1}%
    \setlength\tabcolsep{0pt}%
    \put(0,0){\includegraphics[width=\unitlength,page=1]{2Dplot.pdf}}%
    \put(0.10982143,0.03154768){\makebox(0,0)[lt]{\lineheight{1.25}\smash{\begin{tabular}[t]{l}-1.5\end{tabular}}}}%
    \put(0.24880946,0.03154768){\makebox(0,0)[lt]{\lineheight{1.25}\smash{\begin{tabular}[t]{l}-1\end{tabular}}}}%
    \put(0.36815482,0.03154768){\makebox(0,0)[lt]{\lineheight{1.25}\smash{\begin{tabular}[t]{l}-0.5\end{tabular}}}}%
    \put(0.51071429,0.03154768){\makebox(0,0)[lt]{\lineheight{1.25}\smash{\begin{tabular}[t]{l}0\end{tabular}}}}%
    \put(0.63005946,0.03154768){\makebox(0,0)[lt]{\lineheight{1.25}\smash{\begin{tabular}[t]{l}0.5\end{tabular}}}}%
    \put(0.76904768,0.03154768){\makebox(0,0)[lt]{\lineheight{1.25}\smash{\begin{tabular}[t]{l}1\end{tabular}}}}%
    \put(0.88839286,0.03154768){\makebox(0,0)[lt]{\lineheight{1.25}\smash{\begin{tabular}[t]{l}1.5\end{tabular}}}}%
    \put(0.48392857,-0.00357143){\makebox(0,0)[lt]{\lineheight{1.25}\smash{\begin{tabular}[t]{l}$p_1$[m]\end{tabular}}}}%
    \put(0,0){\includegraphics[width=\unitlength,page=2]{2Dplot.pdf}}%
    \put(0.01904768,0.07767857){\makebox(0,0)[lt]{\lineheight{1.25}\smash{\begin{tabular}[t]{l}-1.5\end{tabular}}}}%
    \put(0.03869054,0.17886911){\makebox(0,0)[lt]{\lineheight{1.25}\smash{\begin{tabular}[t]{l}-1\end{tabular}}}}%
    \put(0.01904768,0.28005946){\makebox(0,0)[lt]{\lineheight{1.25}\smash{\begin{tabular}[t]{l}-0.5\end{tabular}}}}%
    \put(0.05297625,0.38125){\makebox(0,0)[lt]{\lineheight{1.25}\smash{\begin{tabular}[t]{l}0\end{tabular}}}}%
    \put(0.03333339,0.48244054){\makebox(0,0)[lt]{\lineheight{1.25}\smash{\begin{tabular}[t]{l}0.5\end{tabular}}}}%
    \put(0.05297625,0.58363089){\makebox(0,0)[lt]{\lineheight{1.25}\smash{\begin{tabular}[t]{l}1\end{tabular}}}}%
    \put(0.03333339,0.68482143){\makebox(0,0)[lt]{\lineheight{1.25}\smash{\begin{tabular}[t]{l}1.5\end{tabular}}}}%
    \put(0.00178571,0.35714286){\rotatebox{90}{\makebox(0,0)[lt]{\lineheight{1.25}\smash{\begin{tabular}[t]{l}$p_2$[m]\end{tabular}}}}}%
    \put(0,0){\includegraphics[width=\unitlength,page=3]{2Dplot.pdf}}%
    \put(0.55284413,0.6467281){\makebox(0,0)[lt]{\lineheight{1.25}\smash{\begin{tabular}[t]{l}$Y(1)$\end{tabular}}}}%
    \put(0.75414734,0.17624256){\color[rgb]{0,0,0}\makebox(0,0)[lt]{\lineheight{1.25}\smash{\begin{tabular}[t]{l}$Y(2)$\end{tabular}}}}%
    \put(0.80388587,0.30953312){\color[rgb]{0,0,0}\makebox(0,0)[lt]{\lineheight{1.25}\smash{\begin{tabular}[t]{l}$y_{t_2}$\end{tabular}}}}%
    \put(0.32634152,0.50498592){\makebox(0,0)[lt]{\lineheight{1.25}\smash{\begin{tabular}[t]{l}$y_{t_1}$\end{tabular}}}}%
  \end{picture}%
\endgroup%

%% file: images/ballandplate_final/jointplot.pdf_tex
\begingroup%
  \makeatletter%
  \providecommand\color[2][]{%
    \errmessage{(Inkscape) Color is used for the text in Inkscape, but the package 'color.sty' is not loaded}%
    \renewcommand\color[2][]{}%
  }%
  \providecommand\transparent[1]{%
    \errmessage{(Inkscape) Transparency is used (non-zero) for the text in Inkscape, but the package 'transparent.sty' is not loaded}%
    \renewcommand\transparent[1]{}%
  }%
  \providecommand\rotatebox[2]{#2}%
  \newcommand*\fsize{\dimexpr\f@size pt\relax}%
  \newcommand*\lineheight[1]{\fontsize{\fsize}{#1\fsize}\selectfont}%
  \ifx\svgwidth\undefined%
    \setlength{\unitlength}{1385.00170898bp}%
    \ifx\svgscale\undefined%
      \relax%
    \else%
      \setlength{\unitlength}{\unitlength * \real{\svgscale}}%
    \fi%
  \else%
    \setlength{\unitlength}{\svgwidth}%
  \fi%
  \global\let\svgwidth\undefined%
  \global\let\svgscale\undefined%
  \makeatother%
  \begin{picture}(1,0.31219936)%
    \lineheight{1}%
    \setlength\tabcolsep{0pt}%
    \put(0,0){\includegraphics[width=\unitlength,page=1]{jointplot.pdf}}%
    \put(0.04205292,0.17616875){\color[rgb]{0.14901961,0.14901961,0.14901961}\makebox(0,0)[lt]{\lineheight{1.25}\smash{\begin{tabular}[t]{l}0\end{tabular}}}}%
    \put(0.07369775,0.17616875){\color[rgb]{0.14901961,0.14901961,0.14901961}\makebox(0,0)[lt]{\lineheight{1.25}\smash{\begin{tabular}[t]{l}5\end{tabular}}}}%
    \put(0.10344726,0.17616875){\color[rgb]{0.14901961,0.14901961,0.14901961}\makebox(0,0)[lt]{\lineheight{1.25}\smash{\begin{tabular}[t]{l}10\end{tabular}}}}%
    \put(0.13509208,0.17616875){\color[rgb]{0.14901961,0.14901961,0.14901961}\makebox(0,0)[lt]{\lineheight{1.25}\smash{\begin{tabular}[t]{l}15\end{tabular}}}}%
    \put(0,0){\includegraphics[width=\unitlength,page=2]{jointplot.pdf}}%
    \put(0.02947173,0.19010374){\color[rgb]{0.14901961,0.14901961,0.14901961}\makebox(0,0)[lt]{\lineheight{1.25}\smash{\begin{tabular}[t]{l}0\end{tabular}}}}%
    \put(0.01647536,0.22909286){\color[rgb]{0.14901961,0.14901961,0.14901961}\makebox(0,0)[lt]{\lineheight{1.25}\smash{\begin{tabular}[t]{l}0.5\end{tabular}}}}%
    \put(0.02947173,0.26808198){\color[rgb]{0.14901961,0.14901961,0.14901961}\makebox(0,0)[lt]{\lineheight{1.25}\smash{\begin{tabular}[t]{l}1\end{tabular}}}}%
    \put(0.00550061,0.2025586){\color[rgb]{0.14901961,0.14901961,0.14901961}\rotatebox{90}{\makebox(0,0)[lt]{\lineheight{1.25}\smash{\begin{tabular}[t]{l}$\|e_{p_1}\|$[m]\end{tabular}}}}}%
    \put(0,0){\includegraphics[width=\unitlength,page=3]{jointplot.pdf}}%
    \put(0.04205293,0.04674653){\makebox(0,0)[lt]{\lineheight{1.25}\smash{\begin{tabular}[t]{l}0\end{tabular}}}}%
    \put(0.07369775,0.04674653){\makebox(0,0)[lt]{\lineheight{1.25}\smash{\begin{tabular}[t]{l}5\end{tabular}}}}%
    \put(0.10344726,0.04674653){\makebox(0,0)[lt]{\lineheight{1.25}\smash{\begin{tabular}[t]{l}10\end{tabular}}}}%
    \put(0.13509208,0.04674653){\makebox(0,0)[lt]{\lineheight{1.25}\smash{\begin{tabular}[t]{l}15\end{tabular}}}}%
    \put(0.0753562,0.02851549){\makebox(0,0)[lt]{\lineheight{1.25}\smash{\begin{tabular}[t]{l}time[s]\end{tabular}}}}%
    \put(0,0){\includegraphics[width=\unitlength,page=4]{jointplot.pdf}}%
    \put(0.02947174,0.06068151){\makebox(0,0)[lt]{\lineheight{1.25}\smash{\begin{tabular}[t]{l}0\end{tabular}}}}%
    \put(0.01647536,0.0800407){\makebox(0,0)[lt]{\lineheight{1.25}\smash{\begin{tabular}[t]{l}0.5\end{tabular}}}}%
    \put(0.02947174,0.09939988){\makebox(0,0)[lt]{\lineheight{1.25}\smash{\begin{tabular}[t]{l}1\end{tabular}}}}%
    \put(0.01647536,0.11875906){\makebox(0,0)[lt]{\lineheight{1.25}\smash{\begin{tabular}[t]{l}1.5\end{tabular}}}}%
    \put(0.02947174,0.13811824){\makebox(0,0)[lt]{\lineheight{1.25}\smash{\begin{tabular}[t]{l}2\end{tabular}}}}%
    \put(0.00550061,0.07313636){\color[rgb]{0.14901961,0.14901961,0.14901961}\rotatebox{90}{\makebox(0,0)[lt]{\lineheight{1.25}\smash{\begin{tabular}[t]{l}$\|e_{p_2}\|$[m]\end{tabular}}}}}%
    \put(0.05076426,0.30446617){\color[rgb]{0.14901961,0.14901961,0.14901961}\makebox(0,0)[lt]{\lineheight{1.25}\smash{\begin{tabular}[t]{l}$y_{t_1} \in Y_r \cap \tilde{Y}$\end{tabular}}}}%
    \put(0.19254105,0.30488384){\color[rgb]{0.14901961,0.14901961,0.14901961}\makebox(0,0)[lt]{\lineheight{1.25}\smash{\begin{tabular}[t]{l}$y_{t_2} \notin Y_r \cap \tilde{Y}$\end{tabular}}}}%
    \put(0,0){\includegraphics[width=\unitlength,page=5]{jointplot.pdf}}%
    \put(0.18393002,0.17616875){\makebox(0,0)[lt]{\lineheight{1.25}\smash{\begin{tabular}[t]{l}0\end{tabular}}}}%
    \put(0.21574406,0.17616875){\makebox(0,0)[lt]{\lineheight{1.25}\smash{\begin{tabular}[t]{l}5\end{tabular}}}}%
    \put(0.24566279,0.17616875){\makebox(0,0)[lt]{\lineheight{1.25}\smash{\begin{tabular}[t]{l}10\end{tabular}}}}%
    \put(0.27747683,0.17616875){\makebox(0,0)[lt]{\lineheight{1.25}\smash{\begin{tabular}[t]{l}15\end{tabular}}}}%
    \put(0,0){\includegraphics[width=\unitlength,page=6]{jointplot.pdf}}%
    \put(0.17134882,0.19010374){\makebox(0,0)[lt]{\lineheight{1.25}\smash{\begin{tabular}[t]{l}0\end{tabular}}}}%
    \put(0.15835245,0.22908766){\makebox(0,0)[lt]{\lineheight{1.25}\smash{\begin{tabular}[t]{l}0.5\end{tabular}}}}%
    \put(0.17134882,0.26807153){\makebox(0,0)[lt]{\lineheight{1.25}\smash{\begin{tabular}[t]{l}1\end{tabular}}}}%
    \put(0,0){\includegraphics[width=\unitlength,page=7]{jointplot.pdf}}%
    \put(0.18393002,0.04674653){\makebox(0,0)[lt]{\lineheight{1.25}\smash{\begin{tabular}[t]{l}0\end{tabular}}}}%
    \put(0.21574406,0.04674653){\makebox(0,0)[lt]{\lineheight{1.25}\smash{\begin{tabular}[t]{l}5\end{tabular}}}}%
    \put(0.24566279,0.04674653){\makebox(0,0)[lt]{\lineheight{1.25}\smash{\begin{tabular}[t]{l}10\end{tabular}}}}%
    \put(0.27747683,0.04674653){\makebox(0,0)[lt]{\lineheight{1.25}\smash{\begin{tabular}[t]{l}15\end{tabular}}}}%
    \put(0.21750404,0.02851549){\makebox(0,0)[lt]{\lineheight{1.25}\smash{\begin{tabular}[t]{l}time[s]\end{tabular}}}}%
    \put(0,0){\includegraphics[width=\unitlength,page=8]{jointplot.pdf}}%
    \put(0.17134882,0.06068151){\makebox(0,0)[lt]{\lineheight{1.25}\smash{\begin{tabular}[t]{l}0\end{tabular}}}}%
    \put(0.15835245,0.09939988){\makebox(0,0)[lt]{\lineheight{1.25}\smash{\begin{tabular}[t]{l}0.5\end{tabular}}}}%
    \put(0.17134882,0.13811824){\makebox(0,0)[lt]{\lineheight{1.25}\smash{\begin{tabular}[t]{l}1\end{tabular}}}}%
    \put(0,0){\includegraphics[width=\unitlength,page=9]{jointplot.pdf}}%
    \put(0.40904978,0.02833702){\color[rgb]{0,0,0}\makebox(0,0)[lt]{\lineheight{0}\smash{\begin{tabular}[t]{l}time[s]\end{tabular}}}}%
    \put(0.56161961,0.02809561){\color[rgb]{0,0,0}\makebox(0,0)[lt]{\lineheight{1.25}\smash{\begin{tabular}[t]{l}time[s]\end{tabular}}}}%
    \put(0.38837719,0.30556502){\color[rgb]{0.14901961,0.14901961,0.14901961}\makebox(0,0)[lt]{\lineheight{1.25}\smash{\begin{tabular}[t]{l}$y_{t_1} \in Y_r \cap \tilde{Y}$\end{tabular}}}}%
    \put(0.53621188,0.30593917){\color[rgb]{0.14901961,0.14901961,0.14901961}\makebox(0,0)[lt]{\lineheight{1.25}\smash{\begin{tabular}[t]{l}$y_{t_2} \notin Y_r \cap \tilde{Y}$\end{tabular}}}}%
    \put(0.49518287,0.21647903){\color[rgb]{0,0,0}\makebox(0,0)[lt]{\lineheight{1.25}\smash{\begin{tabular}[t]{l}-0.5\end{tabular}}}}%
    \put(0.33367486,0.05953388){\color[rgb]{0.14901961,0.14901961,0.14901961}\rotatebox{90}{\makebox(0,0)[lt]{\lineheight{1.25}\smash{\begin{tabular}[t]{l}$\theta_2$[rad]\end{tabular}}}}}%
    \put(0.46300133,0.04661394){\color[rgb]{0,0,0}\makebox(0,0)[lt]{\lineheight{1.25}\smash{\begin{tabular}[t]{l}15\end{tabular}}}}%
    \put(0.37800623,0.24216581){\color[rgb]{0,0,0}\makebox(0,0)[lt]{\lineheight{1.25}\smash{\begin{tabular}[t]{l}0\end{tabular}}}}%
    \put(0.40709762,0.24208248){\color[rgb]{0,0,0}\makebox(0,0)[lt]{\lineheight{1.25}\smash{\begin{tabular}[t]{l}5\end{tabular}}}}%
    \put(0.43410032,0.24212735){\color[rgb]{0,0,0}\makebox(0,0)[lt]{\lineheight{1.25}\smash{\begin{tabular}[t]{l}10\end{tabular}}}}%
    \put(0.46330345,0.24193516){\color[rgb]{0,0,0}\makebox(0,0)[lt]{\lineheight{1.25}\smash{\begin{tabular}[t]{l}15\end{tabular}}}}%
    \put(0.52900149,0.24178823){\color[rgb]{0,0,0}\makebox(0,0)[lt]{\lineheight{1.25}\smash{\begin{tabular}[t]{l}0\end{tabular}}}}%
    \put(0.55809278,0.2417049){\color[rgb]{0,0,0}\makebox(0,0)[lt]{\lineheight{1.25}\smash{\begin{tabular}[t]{l}5\end{tabular}}}}%
    \put(0.58509531,0.24174976){\color[rgb]{0,0,0}\makebox(0,0)[lt]{\lineheight{1.25}\smash{\begin{tabular}[t]{l}10\end{tabular}}}}%
    \put(0.61429864,0.24155757){\color[rgb]{0,0,0}\makebox(0,0)[lt]{\lineheight{1.25}\smash{\begin{tabular}[t]{l}15\end{tabular}}}}%
    \put(0.52929241,0.17666463){\color[rgb]{0,0,0}\makebox(0,0)[lt]{\lineheight{1.25}\smash{\begin{tabular}[t]{l}0\end{tabular}}}}%
    \put(0.55838369,0.1765813){\color[rgb]{0,0,0}\makebox(0,0)[lt]{\lineheight{1.25}\smash{\begin{tabular}[t]{l}5\end{tabular}}}}%
    \put(0.58538629,0.17662617){\color[rgb]{0,0,0}\makebox(0,0)[lt]{\lineheight{1.25}\smash{\begin{tabular}[t]{l}10\end{tabular}}}}%
    \put(0.61458963,0.17643398){\color[rgb]{0,0,0}\makebox(0,0)[lt]{\lineheight{1.25}\smash{\begin{tabular}[t]{l}15\end{tabular}}}}%
    \put(0.52850091,0.11370712){\color[rgb]{0,0,0}\makebox(0,0)[lt]{\lineheight{1.25}\smash{\begin{tabular}[t]{l}0\end{tabular}}}}%
    \put(0.55759179,0.11362379){\color[rgb]{0,0,0}\makebox(0,0)[lt]{\lineheight{1.25}\smash{\begin{tabular}[t]{l}5\end{tabular}}}}%
    \put(0.58459426,0.11366866){\color[rgb]{0,0,0}\makebox(0,0)[lt]{\lineheight{1.25}\smash{\begin{tabular}[t]{l}10\end{tabular}}}}%
    \put(0.61379759,0.11347645){\color[rgb]{0,0,0}\makebox(0,0)[lt]{\lineheight{1.25}\smash{\begin{tabular}[t]{l}15\end{tabular}}}}%
    \put(0.52929241,0.04592018){\color[rgb]{0,0,0}\makebox(0,0)[lt]{\lineheight{1.25}\smash{\begin{tabular}[t]{l}0\end{tabular}}}}%
    \put(0.55838369,0.04583685){\color[rgb]{0,0,0}\makebox(0,0)[lt]{\lineheight{1.25}\smash{\begin{tabular}[t]{l}5\end{tabular}}}}%
    \put(0.58538629,0.04588172){\color[rgb]{0,0,0}\makebox(0,0)[lt]{\lineheight{1.25}\smash{\begin{tabular}[t]{l}10\end{tabular}}}}%
    \put(0.61458963,0.04568953){\color[rgb]{0,0,0}\makebox(0,0)[lt]{\lineheight{1.25}\smash{\begin{tabular}[t]{l}15\end{tabular}}}}%
    \put(0.3777232,0.17722295){\color[rgb]{0,0,0}\makebox(0,0)[lt]{\lineheight{1.25}\smash{\begin{tabular}[t]{l}0\end{tabular}}}}%
    \put(0.40681408,0.17713962){\color[rgb]{0,0,0}\makebox(0,0)[lt]{\lineheight{1.25}\smash{\begin{tabular}[t]{l}5\end{tabular}}}}%
    \put(0.43381675,0.17718448){\color[rgb]{0,0,0}\makebox(0,0)[lt]{\lineheight{1.25}\smash{\begin{tabular}[t]{l}10\end{tabular}}}}%
    \put(0.46301992,0.17699229){\color[rgb]{0,0,0}\makebox(0,0)[lt]{\lineheight{1.25}\smash{\begin{tabular}[t]{l}15\end{tabular}}}}%
    \put(0.37810969,0.11367052){\color[rgb]{0,0,0}\makebox(0,0)[lt]{\lineheight{1.25}\smash{\begin{tabular}[t]{l}0\end{tabular}}}}%
    \put(0.40720061,0.11358717){\color[rgb]{0,0,0}\makebox(0,0)[lt]{\lineheight{1.25}\smash{\begin{tabular}[t]{l}5\end{tabular}}}}%
    \put(0.43420375,0.11363206){\color[rgb]{0,0,0}\makebox(0,0)[lt]{\lineheight{1.25}\smash{\begin{tabular}[t]{l}10\end{tabular}}}}%
    \put(0.46340691,0.11343985){\color[rgb]{0,0,0}\makebox(0,0)[lt]{\lineheight{1.25}\smash{\begin{tabular}[t]{l}15\end{tabular}}}}%
    \put(0.37770415,0.04684459){\color[rgb]{0,0,0}\makebox(0,0)[lt]{\lineheight{1.25}\smash{\begin{tabular}[t]{l}0\end{tabular}}}}%
    \put(0.40679553,0.04676126){\color[rgb]{0,0,0}\makebox(0,0)[lt]{\lineheight{1.25}\smash{\begin{tabular}[t]{l}5\end{tabular}}}}%
    \put(0.43379817,0.04680613){\color[rgb]{0,0,0}\makebox(0,0)[lt]{\lineheight{1.25}\smash{\begin{tabular}[t]{l}10\end{tabular}}}}%
    \put(0.36354949,0.2859175){\color[rgb]{0,0,0}\makebox(0,0)[lt]{\lineheight{1.25}\smash{\begin{tabular}[t]{l}0\end{tabular}}}}%
    \put(0.36606797,0.1436987){\color[rgb]{0,0,0}\makebox(0,0)[lt]{\lineheight{1.25}\smash{\begin{tabular}[t]{l}0\end{tabular}}}}%
    \put(0.3644622,0.07423926){\color[rgb]{0,0,0}\makebox(0,0)[lt]{\lineheight{1.25}\smash{\begin{tabular}[t]{l}0\end{tabular}}}}%
    \put(0.34499284,0.26925569){\color[rgb]{0,0,0}\makebox(0,0)[lt]{\lineheight{1.25}\smash{\begin{tabular}[t]{l}-0.5\end{tabular}}}}%
    \put(0.35073861,0.22026727){\color[rgb]{0,0,0}\makebox(0,0)[lt]{\lineheight{1.25}\smash{\begin{tabular}[t]{l}0.5\end{tabular}}}}%
    \put(0.35209279,0.09165446){\color[rgb]{0,0,0}\makebox(0,0)[lt]{\lineheight{1.25}\smash{\begin{tabular}[t]{l}0.1\end{tabular}}}}%
    \put(0.34618156,0.05845498){\color[rgb]{0,0,0}\makebox(0,0)[lt]{\lineheight{1.25}\smash{\begin{tabular}[t]{l}-0.1\end{tabular}}}}%
    \put(0.51642135,0.07378343){\color[rgb]{0,0,0}\makebox(0,0)[lt]{\lineheight{1.25}\smash{\begin{tabular}[t]{l}0\end{tabular}}}}%
    \put(0.49489014,0.09119862){\color[rgb]{0,0,0}\makebox(0,0)[lt]{\lineheight{1.25}\smash{\begin{tabular}[t]{l}0.05\end{tabular}}}}%
    \put(0.48897895,0.05799915){\color[rgb]{0,0,0}\makebox(0,0)[lt]{\lineheight{1.25}\smash{\begin{tabular}[t]{l}-0.05\end{tabular}}}}%
    \put(0.51472113,0.28522281){\color[rgb]{0,0,0}\makebox(0,0)[lt]{\lineheight{1.25}\smash{\begin{tabular}[t]{l}  1\end{tabular}}}}%
    \put(0.51505734,0.25396371){\color[rgb]{0,0,0}\makebox(0,0)[lt]{\lineheight{1.25}\smash{\begin{tabular}[t]{l}0\end{tabular}}}}%
    \put(0.34479139,0.1582513){\color[rgb]{0,0,0}\makebox(0,0)[lt]{\lineheight{1.25}\smash{\begin{tabular}[t]{l}0.02\end{tabular}}}}%
    \put(0.33950738,0.12900861){\color[rgb]{0,0,0}\makebox(0,0)[lt]{\lineheight{1.25}\smash{\begin{tabular}[t]{l}-0.02\end{tabular}}}}%
    \put(0.51594129,0.13512182){\color[rgb]{0,0,0}\makebox(0,0)[lt]{\lineheight{1.25}\smash{\begin{tabular}[t]{l}0\end{tabular}}}}%
    \put(0.49466471,0.14967444){\color[rgb]{0,0,0}\makebox(0,0)[lt]{\lineheight{1.25}\smash{\begin{tabular}[t]{l}0.02\end{tabular}}}}%
    \put(0.48829768,0.12043171){\color[rgb]{0,0,0}\makebox(0,0)[lt]{\lineheight{1.25}\smash{\begin{tabular}[t]{l}-0.02\end{tabular}}}}%
    \put(0.35928359,0.25368714){\color[rgb]{0,0,0}\makebox(0,0)[lt]{\lineheight{1.25}\smash{\begin{tabular}[t]{l}-1\end{tabular}}}}%
    \put(0.35923136,0.19163311){\color[rgb]{0,0,0}\makebox(0,0)[lt]{\lineheight{1.25}\smash{\begin{tabular}[t]{l}-1\end{tabular}}}}%
    \put(0,0){\includegraphics[width=\unitlength,page=10]{jointplot.pdf}}%
    \put(0.33344402,0.19176081){\color[rgb]{0.14901961,0.14901961,0.14901961}\rotatebox{90}{\makebox(0,0)[lt]{\lineheight{1.25}\smash{\begin{tabular}[t]{l}$p_2$[m]\end{tabular}}}}}%
    \put(0,0){\includegraphics[width=\unitlength,page=11]{jointplot.pdf}}%
    \put(0.33303898,0.12142342){\color[rgb]{0.14901961,0.14901961,0.14901961}\rotatebox{90}{\makebox(0,0)[lt]{\lineheight{1.25}\smash{\begin{tabular}[t]{l}$\theta_1$[rad]\end{tabular}}}}}%
    \put(0,0){\includegraphics[width=\unitlength,page=12]{jointplot.pdf}}%
    \put(0.33264269,0.25525884){\color[rgb]{0.14901961,0.14901961,0.14901961}\rotatebox{90}{\makebox(0,0)[lt]{\lineheight{1.25}\smash{\begin{tabular}[t]{l}$p_1$[m]\end{tabular}}}}}%
    \put(0.50226726,0.26974797){\color[rgb]{0,0,0}\makebox(0,0)[lt]{\lineheight{1.25}\smash{\begin{tabular}[t]{l}0.5\end{tabular}}}}%
    \put(0.50886502,0.19629753){\color[rgb]{0,0,0}\makebox(0,0)[lt]{\lineheight{1.25}\smash{\begin{tabular}[t]{l}-1\end{tabular}}}}%
    \put(0.67074663,0.19968127){\color[rgb]{0.14901961,0.14901961,0.14901961}\rotatebox{90}{\makebox(0,0)[lt]{\lineheight{1.25}\smash{\begin{tabular}[t]{l}$a_1$[rad$^2$/s]\end{tabular}}}}}%
    \put(0,0){\includegraphics[width=\unitlength,page=13]{jointplot.pdf}}%
    \put(0.71103613,0.17725178){\color[rgb]{0.14901961,0.14901961,0.14901961}\makebox(0,0)[lt]{\lineheight{1.25}\smash{\begin{tabular}[t]{l}0\end{tabular}}}}%
    \put(0.74268095,0.17725178){\color[rgb]{0.14901961,0.14901961,0.14901961}\makebox(0,0)[lt]{\lineheight{1.25}\smash{\begin{tabular}[t]{l}5\end{tabular}}}}%
    \put(0.77243046,0.17725178){\color[rgb]{0.14901961,0.14901961,0.14901961}\makebox(0,0)[lt]{\lineheight{1.25}\smash{\begin{tabular}[t]{l}10\end{tabular}}}}%
    \put(0.80407528,0.17725178){\color[rgb]{0.14901961,0.14901961,0.14901961}\makebox(0,0)[lt]{\lineheight{1.25}\smash{\begin{tabular}[t]{l}15\end{tabular}}}}%
    \put(0,0){\includegraphics[width=\unitlength,page=14]{jointplot.pdf}}%
    \put(0.67787743,0.18902071){\color[rgb]{0.14901961,0.14901961,0.14901961}\makebox(0,0)[lt]{\lineheight{1.25}\smash{\begin{tabular}[t]{l}-0.2\end{tabular}}}}%
    \put(0.67787743,0.20851527){\color[rgb]{0.14901961,0.14901961,0.14901961}\makebox(0,0)[lt]{\lineheight{1.25}\smash{\begin{tabular}[t]{l}-0.1\end{tabular}}}}%
    \put(0.69520592,0.22800983){\color[rgb]{0.14901961,0.14901961,0.14901961}\makebox(0,0)[lt]{\lineheight{1.25}\smash{\begin{tabular}[t]{l}0\end{tabular}}}}%
    \put(0.68437561,0.24750439){\color[rgb]{0.14901961,0.14901961,0.14901961}\makebox(0,0)[lt]{\lineheight{1.25}\smash{\begin{tabular}[t]{l}0.1\end{tabular}}}}%
    \put(0.68437561,0.26699895){\color[rgb]{0.14901961,0.14901961,0.14901961}\makebox(0,0)[lt]{\lineheight{1.25}\smash{\begin{tabular}[t]{l}0.2\end{tabular}}}}%
    \put(0,0){\includegraphics[width=\unitlength,page=15]{jointplot.pdf}}%
    \put(0.71103613,0.0456635){\color[rgb]{0.14901961,0.14901961,0.14901961}\makebox(0,0)[lt]{\lineheight{1.25}\smash{\begin{tabular}[t]{l}0\end{tabular}}}}%
    \put(0.74268095,0.0456635){\color[rgb]{0.14901961,0.14901961,0.14901961}\makebox(0,0)[lt]{\lineheight{1.25}\smash{\begin{tabular}[t]{l}5\end{tabular}}}}%
    \put(0.77243046,0.0456635){\color[rgb]{0.14901961,0.14901961,0.14901961}\makebox(0,0)[lt]{\lineheight{1.25}\smash{\begin{tabular}[t]{l}10\end{tabular}}}}%
    \put(0.80407528,0.0456635){\color[rgb]{0.14901961,0.14901961,0.14901961}\makebox(0,0)[lt]{\lineheight{1.25}\smash{\begin{tabular}[t]{l}15\end{tabular}}}}%
    \put(0.7454225,0.02743246){\color[rgb]{0.14901961,0.14901961,0.14901961}\makebox(0,0)[lt]{\lineheight{1.25}\smash{\begin{tabular}[t]{l}time[s]\end{tabular}}}}%
    \put(0,0){\includegraphics[width=\unitlength,page=16]{jointplot.pdf}}%
    \put(0.67787743,0.0606815){\color[rgb]{0.14901961,0.14901961,0.14901961}\makebox(0,0)[lt]{\lineheight{1.25}\smash{\begin{tabular}[t]{l}-0.2\end{tabular}}}}%
    \put(0.67787743,0.08004068){\color[rgb]{0.14901961,0.14901961,0.14901961}\makebox(0,0)[lt]{\lineheight{1.25}\smash{\begin{tabular}[t]{l}-0.1\end{tabular}}}}%
    \put(0.69520592,0.09939986){\color[rgb]{0.14901961,0.14901961,0.14901961}\makebox(0,0)[lt]{\lineheight{1.25}\smash{\begin{tabular}[t]{l}0\end{tabular}}}}%
    \put(0.68437561,0.11875904){\color[rgb]{0.14901961,0.14901961,0.14901961}\makebox(0,0)[lt]{\lineheight{1.25}\smash{\begin{tabular}[t]{l}0.1\end{tabular}}}}%
    \put(0.68437561,0.13811823){\color[rgb]{0.14901961,0.14901961,0.14901961}\makebox(0,0)[lt]{\lineheight{1.25}\smash{\begin{tabular}[t]{l}0.2\end{tabular}}}}%
    \put(0.67069322,0.07151181){\color[rgb]{0.14901961,0.14901961,0.14901961}\rotatebox{90}{\makebox(0,0)[lt]{\lineheight{1.25}\smash{\begin{tabular}[t]{l}$a_2$[rad$^2$/s]\end{tabular}}}}}%
    \put(0,0){\includegraphics[width=\unitlength,page=17]{jointplot.pdf}}%
    \put(0.8594114,0.17725178){\color[rgb]{0.14901961,0.14901961,0.14901961}\makebox(0,0)[lt]{\lineheight{1.25}\smash{\begin{tabular}[t]{l}0\end{tabular}}}}%
    \put(0.89122544,0.17725178){\color[rgb]{0.14901961,0.14901961,0.14901961}\makebox(0,0)[lt]{\lineheight{1.25}\smash{\begin{tabular}[t]{l}5\end{tabular}}}}%
    \put(0.92114418,0.17725178){\color[rgb]{0.14901961,0.14901961,0.14901961}\makebox(0,0)[lt]{\lineheight{1.25}\smash{\begin{tabular}[t]{l}10\end{tabular}}}}%
    \put(0.95295822,0.17725178){\color[rgb]{0.14901961,0.14901961,0.14901961}\makebox(0,0)[lt]{\lineheight{1.25}\smash{\begin{tabular}[t]{l}15\end{tabular}}}}%
    \put(0,0){\includegraphics[width=\unitlength,page=18]{jointplot.pdf}}%
    \put(0.82625269,0.18902071){\color[rgb]{0.14901961,0.14901961,0.14901961}\makebox(0,0)[lt]{\lineheight{1.25}\smash{\begin{tabular}[t]{l}-0.2\end{tabular}}}}%
    \put(0.82625269,0.20851527){\color[rgb]{0.14901961,0.14901961,0.14901961}\makebox(0,0)[lt]{\lineheight{1.25}\smash{\begin{tabular}[t]{l}-0.1\end{tabular}}}}%
    \put(0.84358119,0.22800983){\color[rgb]{0.14901961,0.14901961,0.14901961}\makebox(0,0)[lt]{\lineheight{1.25}\smash{\begin{tabular}[t]{l}0\end{tabular}}}}%
    \put(0.83275088,0.24750439){\color[rgb]{0.14901961,0.14901961,0.14901961}\makebox(0,0)[lt]{\lineheight{1.25}\smash{\begin{tabular}[t]{l}0.1\end{tabular}}}}%
    \put(0.83275088,0.26699895){\color[rgb]{0.14901961,0.14901961,0.14901961}\makebox(0,0)[lt]{\lineheight{1.25}\smash{\begin{tabular}[t]{l}0.2\end{tabular}}}}%
    \put(0,0){\includegraphics[width=\unitlength,page=19]{jointplot.pdf}}%
    \put(0.8594114,0.0456635){\color[rgb]{0.14901961,0.14901961,0.14901961}\makebox(0,0)[lt]{\lineheight{1.25}\smash{\begin{tabular}[t]{l}0\end{tabular}}}}%
    \put(0.89122544,0.0456635){\color[rgb]{0.14901961,0.14901961,0.14901961}\makebox(0,0)[lt]{\lineheight{1.25}\smash{\begin{tabular}[t]{l}5\end{tabular}}}}%
    \put(0.92114418,0.0456635){\color[rgb]{0.14901961,0.14901961,0.14901961}\makebox(0,0)[lt]{\lineheight{1.25}\smash{\begin{tabular}[t]{l}10\end{tabular}}}}%
    \put(0.95295822,0.0456635){\color[rgb]{0.14901961,0.14901961,0.14901961}\makebox(0,0)[lt]{\lineheight{1.25}\smash{\begin{tabular}[t]{l}15\end{tabular}}}}%
    \put(0.89406853,0.02743246){\color[rgb]{0.14901961,0.14901961,0.14901961}\makebox(0,0)[lt]{\lineheight{1.25}\smash{\begin{tabular}[t]{l}time[s]\end{tabular}}}}%
    \put(0,0){\includegraphics[width=\unitlength,page=20]{jointplot.pdf}}%
    \put(0.82625269,0.0606815){\color[rgb]{0.14901961,0.14901961,0.14901961}\makebox(0,0)[lt]{\lineheight{1.25}\smash{\begin{tabular}[t]{l}-0.2\end{tabular}}}}%
    \put(0.82625269,0.08004068){\color[rgb]{0.14901961,0.14901961,0.14901961}\makebox(0,0)[lt]{\lineheight{1.25}\smash{\begin{tabular}[t]{l}-0.1\end{tabular}}}}%
    \put(0.84358119,0.09939986){\color[rgb]{0.14901961,0.14901961,0.14901961}\makebox(0,0)[lt]{\lineheight{1.25}\smash{\begin{tabular}[t]{l}0\end{tabular}}}}%
    \put(0.83275088,0.11875904){\color[rgb]{0.14901961,0.14901961,0.14901961}\makebox(0,0)[lt]{\lineheight{1.25}\smash{\begin{tabular}[t]{l}0.1\end{tabular}}}}%
    \put(0.83275088,0.13811823){\color[rgb]{0.14901961,0.14901961,0.14901961}\makebox(0,0)[lt]{\lineheight{1.25}\smash{\begin{tabular}[t]{l}0.2\end{tabular}}}}%
    \put(0,0){\includegraphics[width=\unitlength,page=21]{jointplot.pdf}}%
    \put(0.72045792,0.30595477){\color[rgb]{0.14901961,0.14901961,0.14901961}\makebox(0,0)[lt]{\lineheight{1.25}\smash{\begin{tabular}[t]{l}$y_{t_1} \in Y_r \cap \tilde{Y}$\end{tabular}}}}%
    \put(0.86873285,0.30637244){\color[rgb]{0.14901961,0.14901961,0.14901961}\makebox(0,0)[lt]{\lineheight{1.25}\smash{\begin{tabular}[t]{l}$y_{t_2} \notin Y_r \cap \tilde{Y}$\end{tabular}}}}%
    \put(0.1074358,0.00272135){\makebox(0,0)[lt]{\lineheight{1.25}\smash{\begin{tabular}[t]{l}(a) Tracking error\end{tabular}}}}%
    \put(0.46649148,0.00236734){\color[rgb]{0,0,0}\makebox(0,0)[lt]{\lineheight{1.25}\smash{\begin{tabular}[t]{l}(b) States\end{tabular}}}}%
    \put(0.81236776,0.00172326){\color[rgb]{0,0,0}\makebox(0,0)[lt]{\lineheight{1.25}\smash{\begin{tabular}[t]{l}(c) Inputs\end{tabular}}}}%
  \end{picture}%
\endgroup%

%% file: images/quad_final/obstacles.pdf_tex
\begingroup%
  \makeatletter%
  \providecommand\color[2][]{%
    \errmessage{(Inkscape) Color is used for the text in Inkscape, but the package 'color.sty' is not loaded}%
    \renewcommand\color[2][]{}%
  }%
  \providecommand\transparent[1]{%
    \errmessage{(Inkscape) Transparency is used (non-zero) for the text in Inkscape, but the package 'transparent.sty' is not loaded}%
    \renewcommand\transparent[1]{}%
  }%
  \providecommand\rotatebox[2]{#2}%
  \newcommand*\fsize{\dimexpr\f@size pt\relax}%
  \newcommand*\lineheight[1]{\fontsize{\fsize}{#1\fsize}\selectfont}%
  \ifx\svgwidth\undefined%
    \setlength{\unitlength}{356.14512634bp}%
    \ifx\svgscale\undefined%
      \relax%
    \else%
      \setlength{\unitlength}{\unitlength * \real{\svgscale}}%
    \fi%
  \else%
    \setlength{\unitlength}{\svgwidth}%
  \fi%
  \global\let\svgwidth\undefined%
  \global\let\svgscale\undefined%
  \makeatother%
  \begin{picture}(1,0.27916498)%
    \lineheight{1}%
    \setlength\tabcolsep{0pt}%
    \put(0,0){\includegraphics[width=\unitlength,page=1]{obstacles.pdf}}%
    \put(0.07797414,0.04784485){\color[rgb]{0.14901961,0.14901961,0.14901961}\makebox(0,0)[lt]{\lineheight{1.25}\smash{\begin{tabular}[t]{l}0\end{tabular}}}}%
    \put(0.18255967,0.04784485){\color[rgb]{0.14901961,0.14901961,0.14901961}\makebox(0,0)[lt]{\lineheight{1.25}\smash{\begin{tabular}[t]{l}10\end{tabular}}}}%
    \put(0.29346282,0.04784485){\color[rgb]{0.14901961,0.14901961,0.14901961}\makebox(0,0)[lt]{\lineheight{1.25}\smash{\begin{tabular}[t]{l}20\end{tabular}}}}%
    \put(0.40436599,0.04784485){\color[rgb]{0.14901961,0.14901961,0.14901961}\makebox(0,0)[lt]{\lineheight{1.25}\smash{\begin{tabular}[t]{l}30\end{tabular}}}}%
    \put(0.51526942,0.04784485){\color[rgb]{0.14901961,0.14901961,0.14901961}\makebox(0,0)[lt]{\lineheight{1.25}\smash{\begin{tabular}[t]{l}40\end{tabular}}}}%
    \put(0.62617259,0.04784485){\color[rgb]{0.14901961,0.14901961,0.14901961}\makebox(0,0)[lt]{\lineheight{1.25}\smash{\begin{tabular}[t]{l}50\end{tabular}}}}%
    \put(0.73707579,0.04784485){\color[rgb]{0.14901961,0.14901961,0.14901961}\makebox(0,0)[lt]{\lineheight{1.25}\smash{\begin{tabular}[t]{l}60\end{tabular}}}}%
    \put(0.84797893,0.04784485){\color[rgb]{0.14901961,0.14901961,0.14901961}\makebox(0,0)[lt]{\lineheight{1.25}\smash{\begin{tabular}[t]{l}70\end{tabular}}}}%
    \put(0.95888213,0.04784485){\color[rgb]{0.14901961,0.14901961,0.14901961}\makebox(0,0)[lt]{\lineheight{1.25}\smash{\begin{tabular}[t]{l}80\end{tabular}}}}%
    \put(0.49072767,0.00502506){\color[rgb]{0.14901961,0.14901961,0.14901961}\makebox(0,0)[lt]{\lineheight{1.25}\smash{\begin{tabular}[t]{l}time[s]\end{tabular}}}}%
    \put(0,0){\includegraphics[width=\unitlength,page=2]{obstacles.pdf}}%
    \put(0.04336754,0.09508673){\color[rgb]{0.14901961,0.14901961,0.14901961}\makebox(0,0)[lt]{\lineheight{1.25}\smash{\begin{tabular}[t]{l}0\end{tabular}}}}%
    \put(0.04336754,0.1782691){\color[rgb]{0.14901961,0.14901961,0.14901961}\makebox(0,0)[lt]{\lineheight{1.25}\smash{\begin{tabular}[t]{l}1\end{tabular}}}}%
    \put(0.04336754,0.26145149){\color[rgb]{0.14901961,0.14901961,0.14901961}\makebox(0,0)[lt]{\lineheight{1.25}\smash{\begin{tabular}[t]{l}2\end{tabular}}}}%
    \put(0.02020282,0.15405145){\color[rgb]{0.14901961,0.14901961,0.14901961}\rotatebox{90}{\makebox(0,0)[lt]{\lineheight{1.25}\smash{\begin{tabular}[t]{l}$N_o$\end{tabular}}}}}%
    \put(0,0){\includegraphics[width=\unitlength,page=3]{obstacles.pdf}}%
  \end{picture}%
\endgroup%

%% file: images/quad_final/joint.pdf_tex
\begingroup%
  \makeatletter%
  \providecommand\color[2][]{%
    \errmessage{(Inkscape) Color is used for the text in Inkscape, but the package 'color.sty' is not loaded}%
    \renewcommand\color[2][]{}%
  }%
  \providecommand\transparent[1]{%
    \errmessage{(Inkscape) Transparency is used (non-zero) for the text in Inkscape, but the package 'transparent.sty' is not loaded}%
    \renewcommand\transparent[1]{}%
  }%
  \providecommand\rotatebox[2]{#2}%
  \newcommand*\fsize{\dimexpr\f@size pt\relax}%
  \newcommand*\lineheight[1]{\fontsize{\fsize}{#1\fsize}\selectfont}%
  \ifx\svgwidth\undefined%
    \setlength{\unitlength}{1189.9420166bp}%
    \ifx\svgscale\undefined%
      \relax%
    \else%
      \setlength{\unitlength}{\unitlength * \real{\svgscale}}%
    \fi%
  \else%
    \setlength{\unitlength}{\svgwidth}%
  \fi%
  \global\let\svgwidth\undefined%
  \global\let\svgscale\undefined%
  \makeatother%
  \begin{picture}(1,0.28636492)%
    \lineheight{1}%
    \setlength\tabcolsep{0pt}%
    \put(0.00226493,0.06663919){\color[rgb]{0.14901961,0.14901961,0.14901961}\rotatebox{90}{\makebox(0,0)[lt]{\lineheight{1.25}\smash{\begin{tabular}[t]{l}$\|e_z\|$[m]\end{tabular}}}}}%
    \put(0.00226493,0.14983652){\color[rgb]{0.14901961,0.14901961,0.14901961}\rotatebox{90}{\makebox(0,0)[lt]{\lineheight{1.25}\smash{\begin{tabular}[t]{l}$\|e_y\|$[m]\end{tabular}}}}}%
    \put(0.00226493,0.23303386){\color[rgb]{0.14901961,0.14901961,0.14901961}\rotatebox{90}{\makebox(0,0)[lt]{\lineheight{1.25}\smash{\begin{tabular}[t]{l}$\|e_x\|$[m]\end{tabular}}}}}%
    \put(0,0){\includegraphics[width=\unitlength,page=1]{joint.pdf}}%
    \put(0.02968223,0.21290685){\color[rgb]{0.14901961,0.14901961,0.14901961}\makebox(0,0)[lt]{\lineheight{1.25}\smash{\begin{tabular}[t]{l}0\end{tabular}}}}%
    \put(0.06098428,0.21290685){\color[rgb]{0.14901961,0.14901961,0.14901961}\makebox(0,0)[lt]{\lineheight{1.25}\smash{\begin{tabular}[t]{l}10\end{tabular}}}}%
    \put(0.09417718,0.21290685){\color[rgb]{0.14901961,0.14901961,0.14901961}\makebox(0,0)[lt]{\lineheight{1.25}\smash{\begin{tabular}[t]{l}20\end{tabular}}}}%
    \put(0.12737008,0.21290685){\color[rgb]{0.14901961,0.14901961,0.14901961}\makebox(0,0)[lt]{\lineheight{1.25}\smash{\begin{tabular}[t]{l}30\end{tabular}}}}%
    \put(0.16056307,0.21290685){\color[rgb]{0.14901961,0.14901961,0.14901961}\makebox(0,0)[lt]{\lineheight{1.25}\smash{\begin{tabular}[t]{l}40\end{tabular}}}}%
    \put(0.19375593,0.21290685){\color[rgb]{0.14901961,0.14901961,0.14901961}\makebox(0,0)[lt]{\lineheight{1.25}\smash{\begin{tabular}[t]{l}50\end{tabular}}}}%
    \put(0.22694884,0.21290685){\color[rgb]{0.14901961,0.14901961,0.14901961}\makebox(0,0)[lt]{\lineheight{1.25}\smash{\begin{tabular}[t]{l}60\end{tabular}}}}%
    \put(0.26014173,0.21290685){\color[rgb]{0.14901961,0.14901961,0.14901961}\makebox(0,0)[lt]{\lineheight{1.25}\smash{\begin{tabular}[t]{l}70\end{tabular}}}}%
    \put(0.29333464,0.21290685){\color[rgb]{0.14901961,0.14901961,0.14901961}\makebox(0,0)[lt]{\lineheight{1.25}\smash{\begin{tabular}[t]{l}80\end{tabular}}}}%
    \put(0,0){\includegraphics[width=\unitlength,page=2]{joint.pdf}}%
    \put(0.01302178,0.22200392){\color[rgb]{0.14901961,0.14901961,0.14901961}\makebox(0,0)[lt]{\lineheight{1.25}\smash{\begin{tabular}[t]{l}0\end{tabular}}}}%
    \put(0.00924008,0.25068176){\color[rgb]{0.14901961,0.14901961,0.14901961}\makebox(0,0)[lt]{\lineheight{1.25}\smash{\begin{tabular}[t]{l}20\end{tabular}}}}%
    \put(0.00924008,0.27935964){\color[rgb]{0.14901961,0.14901961,0.14901961}\makebox(0,0)[lt]{\lineheight{1.25}\smash{\begin{tabular}[t]{l}40\end{tabular}}}}%
    \put(0,0){\includegraphics[width=\unitlength,page=3]{joint.pdf}}%
    \put(0.02968223,0.12844896){\color[rgb]{0.14901961,0.14901961,0.14901961}\makebox(0,0)[lt]{\lineheight{1.25}\smash{\begin{tabular}[t]{l}0\end{tabular}}}}%
    \put(0.06098428,0.12844896){\color[rgb]{0.14901961,0.14901961,0.14901961}\makebox(0,0)[lt]{\lineheight{1.25}\smash{\begin{tabular}[t]{l}10\end{tabular}}}}%
    \put(0.09417718,0.12844896){\color[rgb]{0.14901961,0.14901961,0.14901961}\makebox(0,0)[lt]{\lineheight{1.25}\smash{\begin{tabular}[t]{l}20\end{tabular}}}}%
    \put(0.12737008,0.12844896){\color[rgb]{0.14901961,0.14901961,0.14901961}\makebox(0,0)[lt]{\lineheight{1.25}\smash{\begin{tabular}[t]{l}30\end{tabular}}}}%
    \put(0.16056307,0.12844896){\color[rgb]{0.14901961,0.14901961,0.14901961}\makebox(0,0)[lt]{\lineheight{1.25}\smash{\begin{tabular}[t]{l}40\end{tabular}}}}%
    \put(0.19375593,0.12844896){\color[rgb]{0.14901961,0.14901961,0.14901961}\makebox(0,0)[lt]{\lineheight{1.25}\smash{\begin{tabular}[t]{l}50\end{tabular}}}}%
    \put(0.22694884,0.12844896){\color[rgb]{0.14901961,0.14901961,0.14901961}\makebox(0,0)[lt]{\lineheight{1.25}\smash{\begin{tabular}[t]{l}60\end{tabular}}}}%
    \put(0.26014173,0.12844896){\color[rgb]{0.14901961,0.14901961,0.14901961}\makebox(0,0)[lt]{\lineheight{1.25}\smash{\begin{tabular}[t]{l}70\end{tabular}}}}%
    \put(0.29333464,0.12844896){\color[rgb]{0.14901961,0.14901961,0.14901961}\makebox(0,0)[lt]{\lineheight{1.25}\smash{\begin{tabular}[t]{l}80\end{tabular}}}}%
    \put(0,0){\includegraphics[width=\unitlength,page=4]{joint.pdf}}%
    \put(0.01302178,0.13754601){\color[rgb]{0.14901961,0.14901961,0.14901961}\makebox(0,0)[lt]{\lineheight{1.25}\smash{\begin{tabular}[t]{l}0\end{tabular}}}}%
    \put(0.00924008,0.16622388){\color[rgb]{0.14901961,0.14901961,0.14901961}\makebox(0,0)[lt]{\lineheight{1.25}\smash{\begin{tabular}[t]{l}10\end{tabular}}}}%
    \put(0.00924008,0.19490175){\color[rgb]{0.14901961,0.14901961,0.14901961}\makebox(0,0)[lt]{\lineheight{1.25}\smash{\begin{tabular}[t]{l}20\end{tabular}}}}%
    \put(0,0){\includegraphics[width=\unitlength,page=5]{joint.pdf}}%
    \put(0.02968223,0.04399106){\color[rgb]{0.14901961,0.14901961,0.14901961}\makebox(0,0)[lt]{\lineheight{1.25}\smash{\begin{tabular}[t]{l}0\end{tabular}}}}%
    \put(0.06098428,0.04399106){\color[rgb]{0.14901961,0.14901961,0.14901961}\makebox(0,0)[lt]{\lineheight{1.25}\smash{\begin{tabular}[t]{l}10\end{tabular}}}}%
    \put(0.09417718,0.04399106){\color[rgb]{0.14901961,0.14901961,0.14901961}\makebox(0,0)[lt]{\lineheight{1.25}\smash{\begin{tabular}[t]{l}20\end{tabular}}}}%
    \put(0.12737008,0.04399106){\color[rgb]{0.14901961,0.14901961,0.14901961}\makebox(0,0)[lt]{\lineheight{1.25}\smash{\begin{tabular}[t]{l}30\end{tabular}}}}%
    \put(0.16056307,0.04399106){\color[rgb]{0.14901961,0.14901961,0.14901961}\makebox(0,0)[lt]{\lineheight{1.25}\smash{\begin{tabular}[t]{l}40\end{tabular}}}}%
    \put(0.19375593,0.04399106){\color[rgb]{0.14901961,0.14901961,0.14901961}\makebox(0,0)[lt]{\lineheight{1.25}\smash{\begin{tabular}[t]{l}50\end{tabular}}}}%
    \put(0.22694884,0.04399106){\color[rgb]{0.14901961,0.14901961,0.14901961}\makebox(0,0)[lt]{\lineheight{1.25}\smash{\begin{tabular}[t]{l}60\end{tabular}}}}%
    \put(0.26014173,0.04399106){\color[rgb]{0.14901961,0.14901961,0.14901961}\makebox(0,0)[lt]{\lineheight{1.25}\smash{\begin{tabular}[t]{l}70\end{tabular}}}}%
    \put(0.29333464,0.04399106){\color[rgb]{0.14901961,0.14901961,0.14901961}\makebox(0,0)[lt]{\lineheight{1.25}\smash{\begin{tabular}[t]{l}80\end{tabular}}}}%
    \put(0.15195724,0.02739357){\color[rgb]{0.14901961,0.14901961,0.14901961}\makebox(0,0)[lt]{\lineheight{1.25}\smash{\begin{tabular}[t]{l}time[s]\end{tabular}}}}%
    \put(0.11346747,0.00162293){\color[rgb]{0.14901961,0.14901961,0.14901961}\makebox(0,0)[lt]{\lineheight{1.25}\smash{\begin{tabular}[t]{l}(a) Tracking error\end{tabular}}}}%
    \put(0,0){\includegraphics[width=\unitlength,page=6]{joint.pdf}}%
    \put(0.01302178,0.05308811){\color[rgb]{0.14901961,0.14901961,0.14901961}\makebox(0,0)[lt]{\lineheight{1.25}\smash{\begin{tabular}[t]{l}0\end{tabular}}}}%
    \put(0.01302178,0.07220668){\color[rgb]{0.14901961,0.14901961,0.14901961}\makebox(0,0)[lt]{\lineheight{1.25}\smash{\begin{tabular}[t]{l}5\end{tabular}}}}%
    \put(0.00924008,0.09132528){\color[rgb]{0.14901961,0.14901961,0.14901961}\makebox(0,0)[lt]{\lineheight{1.25}\smash{\begin{tabular}[t]{l}10\end{tabular}}}}%
    \put(0.00924008,0.11044385){\color[rgb]{0.14901961,0.14901961,0.14901961}\makebox(0,0)[lt]{\lineheight{1.25}\smash{\begin{tabular}[t]{l}15\end{tabular}}}}%
    \put(0,0){\includegraphics[width=\unitlength,page=7]{joint.pdf}}%
    \put(0.38299294,0.2117623){\color[rgb]{0.14901961,0.14901961,0.14901961}\makebox(0,0)[lt]{\lineheight{1.25}\smash{\begin{tabular}[t]{l}0\end{tabular}}}}%
    \put(0.414295,0.2117623){\color[rgb]{0.14901961,0.14901961,0.14901961}\makebox(0,0)[lt]{\lineheight{1.25}\smash{\begin{tabular}[t]{l}10\end{tabular}}}}%
    \put(0.44748787,0.2117623){\color[rgb]{0.14901961,0.14901961,0.14901961}\makebox(0,0)[lt]{\lineheight{1.25}\smash{\begin{tabular}[t]{l}20\end{tabular}}}}%
    \put(0.48068078,0.2117623){\color[rgb]{0.14901961,0.14901961,0.14901961}\makebox(0,0)[lt]{\lineheight{1.25}\smash{\begin{tabular}[t]{l}30\end{tabular}}}}%
    \put(0.51387376,0.2117623){\color[rgb]{0.14901961,0.14901961,0.14901961}\makebox(0,0)[lt]{\lineheight{1.25}\smash{\begin{tabular}[t]{l}40\end{tabular}}}}%
    \put(0.54706663,0.2117623){\color[rgb]{0.14901961,0.14901961,0.14901961}\makebox(0,0)[lt]{\lineheight{1.25}\smash{\begin{tabular}[t]{l}50\end{tabular}}}}%
    \put(0.58025954,0.2117623){\color[rgb]{0.14901961,0.14901961,0.14901961}\makebox(0,0)[lt]{\lineheight{1.25}\smash{\begin{tabular}[t]{l}60\end{tabular}}}}%
    \put(0.61345244,0.2117623){\color[rgb]{0.14901961,0.14901961,0.14901961}\makebox(0,0)[lt]{\lineheight{1.25}\smash{\begin{tabular}[t]{l}70\end{tabular}}}}%
    \put(0.64664531,0.2117623){\color[rgb]{0.14901961,0.14901961,0.14901961}\makebox(0,0)[lt]{\lineheight{1.25}\smash{\begin{tabular}[t]{l}80\end{tabular}}}}%
    \put(0,0){\includegraphics[width=\unitlength,page=8]{joint.pdf}}%
    \put(0.35183599,0.22338049){\color[rgb]{0.14901961,0.14901961,0.14901961}\makebox(0,0)[lt]{\lineheight{1.25}\smash{\begin{tabular}[t]{l}-0.2\end{tabular}}}}%
    \put(0.36759306,0.25205835){\color[rgb]{0.14901961,0.14901961,0.14901961}\makebox(0,0)[lt]{\lineheight{1.25}\smash{\begin{tabular}[t]{l}0\end{tabular}}}}%
    \put(0.35687825,0.28073622){\color[rgb]{0.14901961,0.14901961,0.14901961}\makebox(0,0)[lt]{\lineheight{1.25}\smash{\begin{tabular}[t]{l}0.2\end{tabular}}}}%
    \put(0.33733949,0.235671){\color[rgb]{0.14901961,0.14901961,0.14901961}\rotatebox{90}{\makebox(0,0)[lt]{\lineheight{1.25}\smash{\begin{tabular}[t]{l}$\phi$[rad]\end{tabular}}}}}%
    \put(0,0){\includegraphics[width=\unitlength,page=9]{joint.pdf}}%
    \put(0.38299294,0.12856498){\color[rgb]{0.14901961,0.14901961,0.14901961}\makebox(0,0)[lt]{\lineheight{1.25}\smash{\begin{tabular}[t]{l}0\end{tabular}}}}%
    \put(0.414295,0.12856498){\color[rgb]{0.14901961,0.14901961,0.14901961}\makebox(0,0)[lt]{\lineheight{1.25}\smash{\begin{tabular}[t]{l}10\end{tabular}}}}%
    \put(0.44748787,0.12856498){\color[rgb]{0.14901961,0.14901961,0.14901961}\makebox(0,0)[lt]{\lineheight{1.25}\smash{\begin{tabular}[t]{l}20\end{tabular}}}}%
    \put(0.48068078,0.12856498){\color[rgb]{0.14901961,0.14901961,0.14901961}\makebox(0,0)[lt]{\lineheight{1.25}\smash{\begin{tabular}[t]{l}30\end{tabular}}}}%
    \put(0.51387376,0.12856498){\color[rgb]{0.14901961,0.14901961,0.14901961}\makebox(0,0)[lt]{\lineheight{1.25}\smash{\begin{tabular}[t]{l}40\end{tabular}}}}%
    \put(0.54706663,0.12856498){\color[rgb]{0.14901961,0.14901961,0.14901961}\makebox(0,0)[lt]{\lineheight{1.25}\smash{\begin{tabular}[t]{l}50\end{tabular}}}}%
    \put(0.58025954,0.12856498){\color[rgb]{0.14901961,0.14901961,0.14901961}\makebox(0,0)[lt]{\lineheight{1.25}\smash{\begin{tabular}[t]{l}60\end{tabular}}}}%
    \put(0.61345244,0.12856498){\color[rgb]{0.14901961,0.14901961,0.14901961}\makebox(0,0)[lt]{\lineheight{1.25}\smash{\begin{tabular}[t]{l}70\end{tabular}}}}%
    \put(0.64664531,0.12856498){\color[rgb]{0.14901961,0.14901961,0.14901961}\makebox(0,0)[lt]{\lineheight{1.25}\smash{\begin{tabular}[t]{l}80\end{tabular}}}}%
    \put(0,0){\includegraphics[width=\unitlength,page=10]{joint.pdf}}%
    \put(0.35183599,0.14201344){\color[rgb]{0.14901961,0.14901961,0.14901961}\makebox(0,0)[lt]{\lineheight{1.25}\smash{\begin{tabular}[t]{l}-0.2\end{tabular}}}}%
    \put(0.36759306,0.1636196){\color[rgb]{0.14901961,0.14901961,0.14901961}\makebox(0,0)[lt]{\lineheight{1.25}\smash{\begin{tabular}[t]{l}0\end{tabular}}}}%
    \put(0.35687825,0.18522576){\color[rgb]{0.14901961,0.14901961,0.14901961}\makebox(0,0)[lt]{\lineheight{1.25}\smash{\begin{tabular}[t]{l}0.2\end{tabular}}}}%
    \put(0.33733949,0.15121311){\color[rgb]{0.14901961,0.14901961,0.14901961}\rotatebox{90}{\makebox(0,0)[lt]{\lineheight{1.25}\smash{\begin{tabular}[t]{l}$\theta$[rad]\end{tabular}}}}}%
    \put(0,0){\includegraphics[width=\unitlength,page=11]{joint.pdf}}%
    \put(0.38299294,0.04410708){\color[rgb]{0.14901961,0.14901961,0.14901961}\makebox(0,0)[lt]{\lineheight{1.25}\smash{\begin{tabular}[t]{l}0\end{tabular}}}}%
    \put(0.414295,0.04410708){\color[rgb]{0.14901961,0.14901961,0.14901961}\makebox(0,0)[lt]{\lineheight{1.25}\smash{\begin{tabular}[t]{l}10\end{tabular}}}}%
    \put(0.44748787,0.04410708){\color[rgb]{0.14901961,0.14901961,0.14901961}\makebox(0,0)[lt]{\lineheight{1.25}\smash{\begin{tabular}[t]{l}20\end{tabular}}}}%
    \put(0.48068078,0.04410708){\color[rgb]{0.14901961,0.14901961,0.14901961}\makebox(0,0)[lt]{\lineheight{1.25}\smash{\begin{tabular}[t]{l}30\end{tabular}}}}%
    \put(0.51387376,0.04410708){\color[rgb]{0.14901961,0.14901961,0.14901961}\makebox(0,0)[lt]{\lineheight{1.25}\smash{\begin{tabular}[t]{l}40\end{tabular}}}}%
    \put(0.54706663,0.04410708){\color[rgb]{0.14901961,0.14901961,0.14901961}\makebox(0,0)[lt]{\lineheight{1.25}\smash{\begin{tabular}[t]{l}50\end{tabular}}}}%
    \put(0.58025954,0.04410708){\color[rgb]{0.14901961,0.14901961,0.14901961}\makebox(0,0)[lt]{\lineheight{1.25}\smash{\begin{tabular}[t]{l}60\end{tabular}}}}%
    \put(0.61345244,0.04410708){\color[rgb]{0.14901961,0.14901961,0.14901961}\makebox(0,0)[lt]{\lineheight{1.25}\smash{\begin{tabular}[t]{l}70\end{tabular}}}}%
    \put(0.64664531,0.04410708){\color[rgb]{0.14901961,0.14901961,0.14901961}\makebox(0,0)[lt]{\lineheight{1.25}\smash{\begin{tabular}[t]{l}80\end{tabular}}}}%
    \put(0.50652849,0.02750959){\color[rgb]{0.14901961,0.14901961,0.14901961}\makebox(0,0)[lt]{\lineheight{1.25}\smash{\begin{tabular}[t]{l}time[s]\end{tabular}}}}%
    \put(0,0){\includegraphics[width=\unitlength,page=12]{joint.pdf}}%
    \put(0.3442726,0.06546074){\color[rgb]{0.14901961,0.14901961,0.14901961}\makebox(0,0)[lt]{\lineheight{1.25}\smash{\begin{tabular}[t]{l}-0.04\end{tabular}}}}%
    \put(0.3442726,0.0806164){\color[rgb]{0.14901961,0.14901961,0.14901961}\makebox(0,0)[lt]{\lineheight{1.25}\smash{\begin{tabular}[t]{l}-0.02\end{tabular}}}}%
    \put(0.36759306,0.09577206){\color[rgb]{0.14901961,0.14901961,0.14901961}\makebox(0,0)[lt]{\lineheight{1.25}\smash{\begin{tabular}[t]{l}0\end{tabular}}}}%
    \put(0.33733949,0.07053691){\color[rgb]{0.14901961,0.14901961,0.14901961}\rotatebox{90}{\makebox(0,0)[lt]{\lineheight{1.25}\smash{\begin{tabular}[t]{l}$\psi$[rad]\end{tabular}}}}}%
    \put(0,0){\includegraphics[width=\unitlength,page=13]{joint.pdf}}%
    \put(0.72281859,0.22872419){\color[rgb]{0.14901961,0.14901961,0.14901961}\makebox(0,0)[lt]{\lineheight{1.25}\smash{\begin{tabular}[t]{l}0\end{tabular}}}}%
    \put(0.75412061,0.22872419){\color[rgb]{0.14901961,0.14901961,0.14901961}\makebox(0,0)[lt]{\lineheight{1.25}\smash{\begin{tabular}[t]{l}10\end{tabular}}}}%
    \put(0.78731356,0.22872419){\color[rgb]{0.14901961,0.14901961,0.14901961}\makebox(0,0)[lt]{\lineheight{1.25}\smash{\begin{tabular}[t]{l}20\end{tabular}}}}%
    \put(0.82050643,0.22872419){\color[rgb]{0.14901961,0.14901961,0.14901961}\makebox(0,0)[lt]{\lineheight{1.25}\smash{\begin{tabular}[t]{l}30\end{tabular}}}}%
    \put(0.85369945,0.22872419){\color[rgb]{0.14901961,0.14901961,0.14901961}\makebox(0,0)[lt]{\lineheight{1.25}\smash{\begin{tabular}[t]{l}40\end{tabular}}}}%
    \put(0.88689232,0.22872419){\color[rgb]{0.14901961,0.14901961,0.14901961}\makebox(0,0)[lt]{\lineheight{1.25}\smash{\begin{tabular}[t]{l}50\end{tabular}}}}%
    \put(0.92008519,0.22872419){\color[rgb]{0.14901961,0.14901961,0.14901961}\makebox(0,0)[lt]{\lineheight{1.25}\smash{\begin{tabular}[t]{l}60\end{tabular}}}}%
    \put(0.95327806,0.22872419){\color[rgb]{0.14901961,0.14901961,0.14901961}\makebox(0,0)[lt]{\lineheight{1.25}\smash{\begin{tabular}[t]{l}70\end{tabular}}}}%
    \put(0.986471,0.22872419){\color[rgb]{0.14901961,0.14901961,0.14901961}\makebox(0,0)[lt]{\lineheight{1.25}\smash{\begin{tabular}[t]{l}80\end{tabular}}}}%
    \put(0,0){\includegraphics[width=\unitlength,page=14]{joint.pdf}}%
    \put(0.7069775,0.24078358){\color[rgb]{0.14901961,0.14901961,0.14901961}\makebox(0,0)[lt]{\lineheight{1.25}\smash{\begin{tabular}[t]{l}0\end{tabular}}}}%
    \put(0.7069775,0.25837899){\color[rgb]{0.14901961,0.14901961,0.14901961}\makebox(0,0)[lt]{\lineheight{1.25}\smash{\begin{tabular}[t]{l}5\end{tabular}}}}%
    \put(0.69941411,0.27597434){\color[rgb]{0.14901961,0.14901961,0.14901961}\makebox(0,0)[lt]{\lineheight{1.25}\smash{\begin{tabular}[t]{l}10\end{tabular}}}}%
    \put(0.6916406,0.24771669){\color[rgb]{0.14901961,0.14901961,0.14901961}\rotatebox{90}{\makebox(0,0)[lt]{\lineheight{1.25}\smash{\begin{tabular}[t]{l}$f_1$[N]\end{tabular}}}}}%
    \put(0,0){\includegraphics[width=\unitlength,page=15]{joint.pdf}}%
    \put(0.72281859,0.16695647){\color[rgb]{0.14901961,0.14901961,0.14901961}\makebox(0,0)[lt]{\lineheight{1.25}\smash{\begin{tabular}[t]{l}0\end{tabular}}}}%
    \put(0.75412061,0.16695647){\color[rgb]{0.14901961,0.14901961,0.14901961}\makebox(0,0)[lt]{\lineheight{1.25}\smash{\begin{tabular}[t]{l}10\end{tabular}}}}%
    \put(0.78731356,0.16695647){\color[rgb]{0.14901961,0.14901961,0.14901961}\makebox(0,0)[lt]{\lineheight{1.25}\smash{\begin{tabular}[t]{l}20\end{tabular}}}}%
    \put(0.82050643,0.16695647){\color[rgb]{0.14901961,0.14901961,0.14901961}\makebox(0,0)[lt]{\lineheight{1.25}\smash{\begin{tabular}[t]{l}30\end{tabular}}}}%
    \put(0.85369945,0.16695647){\color[rgb]{0.14901961,0.14901961,0.14901961}\makebox(0,0)[lt]{\lineheight{1.25}\smash{\begin{tabular}[t]{l}40\end{tabular}}}}%
    \put(0.88689232,0.16695647){\color[rgb]{0.14901961,0.14901961,0.14901961}\makebox(0,0)[lt]{\lineheight{1.25}\smash{\begin{tabular}[t]{l}50\end{tabular}}}}%
    \put(0.92008519,0.16695647){\color[rgb]{0.14901961,0.14901961,0.14901961}\makebox(0,0)[lt]{\lineheight{1.25}\smash{\begin{tabular}[t]{l}60\end{tabular}}}}%
    \put(0.95327806,0.16695647){\color[rgb]{0.14901961,0.14901961,0.14901961}\makebox(0,0)[lt]{\lineheight{1.25}\smash{\begin{tabular}[t]{l}70\end{tabular}}}}%
    \put(0.986471,0.16695647){\color[rgb]{0.14901961,0.14901961,0.14901961}\makebox(0,0)[lt]{\lineheight{1.25}\smash{\begin{tabular}[t]{l}80\end{tabular}}}}%
    \put(0,0){\includegraphics[width=\unitlength,page=16]{joint.pdf}}%
    \put(0.7069775,0.17901586){\color[rgb]{0.14901961,0.14901961,0.14901961}\makebox(0,0)[lt]{\lineheight{1.25}\smash{\begin{tabular}[t]{l}0\end{tabular}}}}%
    \put(0.7069775,0.19634865){\color[rgb]{0.14901961,0.14901961,0.14901961}\makebox(0,0)[lt]{\lineheight{1.25}\smash{\begin{tabular}[t]{l}5\end{tabular}}}}%
    \put(0.69941411,0.21368142){\color[rgb]{0.14901961,0.14901961,0.14901961}\makebox(0,0)[lt]{\lineheight{1.25}\smash{\begin{tabular}[t]{l}10\end{tabular}}}}%
    \put(0.6916406,0.18563383){\color[rgb]{0.14901961,0.14901961,0.14901961}\rotatebox{90}{\makebox(0,0)[lt]{\lineheight{1.25}\smash{\begin{tabular}[t]{l}$f_2$[N]\end{tabular}}}}}%
    \put(0,0){\includegraphics[width=\unitlength,page=17]{joint.pdf}}%
    \put(0.72281859,0.10518875){\color[rgb]{0.14901961,0.14901961,0.14901961}\makebox(0,0)[lt]{\lineheight{1.25}\smash{\begin{tabular}[t]{l}0\end{tabular}}}}%
    \put(0.75412061,0.10518875){\color[rgb]{0.14901961,0.14901961,0.14901961}\makebox(0,0)[lt]{\lineheight{1.25}\smash{\begin{tabular}[t]{l}10\end{tabular}}}}%
    \put(0.78731356,0.10518875){\color[rgb]{0.14901961,0.14901961,0.14901961}\makebox(0,0)[lt]{\lineheight{1.25}\smash{\begin{tabular}[t]{l}20\end{tabular}}}}%
    \put(0.82050643,0.10518875){\color[rgb]{0.14901961,0.14901961,0.14901961}\makebox(0,0)[lt]{\lineheight{1.25}\smash{\begin{tabular}[t]{l}30\end{tabular}}}}%
    \put(0.85369945,0.10518875){\color[rgb]{0.14901961,0.14901961,0.14901961}\makebox(0,0)[lt]{\lineheight{1.25}\smash{\begin{tabular}[t]{l}40\end{tabular}}}}%
    \put(0.88689232,0.10518875){\color[rgb]{0.14901961,0.14901961,0.14901961}\makebox(0,0)[lt]{\lineheight{1.25}\smash{\begin{tabular}[t]{l}50\end{tabular}}}}%
    \put(0.92008519,0.10518875){\color[rgb]{0.14901961,0.14901961,0.14901961}\makebox(0,0)[lt]{\lineheight{1.25}\smash{\begin{tabular}[t]{l}60\end{tabular}}}}%
    \put(0.95327806,0.10518875){\color[rgb]{0.14901961,0.14901961,0.14901961}\makebox(0,0)[lt]{\lineheight{1.25}\smash{\begin{tabular}[t]{l}70\end{tabular}}}}%
    \put(0.986471,0.10518875){\color[rgb]{0.14901961,0.14901961,0.14901961}\makebox(0,0)[lt]{\lineheight{1.25}\smash{\begin{tabular}[t]{l}80\end{tabular}}}}%
    \put(0,0){\includegraphics[width=\unitlength,page=18]{joint.pdf}}%
    \put(0.7069775,0.11724814){\color[rgb]{0.14901961,0.14901961,0.14901961}\makebox(0,0)[lt]{\lineheight{1.25}\smash{\begin{tabular}[t]{l}0\end{tabular}}}}%
    \put(0.7069775,0.13458093){\color[rgb]{0.14901961,0.14901961,0.14901961}\makebox(0,0)[lt]{\lineheight{1.25}\smash{\begin{tabular}[t]{l}5\end{tabular}}}}%
    \put(0.69941411,0.15191371){\color[rgb]{0.14901961,0.14901961,0.14901961}\makebox(0,0)[lt]{\lineheight{1.25}\smash{\begin{tabular}[t]{l}10\end{tabular}}}}%
    \put(0.6916406,0.12386611){\color[rgb]{0.14901961,0.14901961,0.14901961}\rotatebox{90}{\makebox(0,0)[lt]{\lineheight{1.25}\smash{\begin{tabular}[t]{l}$f_3$[N]\end{tabular}}}}}%
    \put(0,0){\includegraphics[width=\unitlength,page=19]{joint.pdf}}%
    \put(0.72281859,0.04342104){\color[rgb]{0.14901961,0.14901961,0.14901961}\makebox(0,0)[lt]{\lineheight{1.25}\smash{\begin{tabular}[t]{l}0\end{tabular}}}}%
    \put(0.75412061,0.04342104){\color[rgb]{0.14901961,0.14901961,0.14901961}\makebox(0,0)[lt]{\lineheight{1.25}\smash{\begin{tabular}[t]{l}10\end{tabular}}}}%
    \put(0.78731356,0.04342104){\color[rgb]{0.14901961,0.14901961,0.14901961}\makebox(0,0)[lt]{\lineheight{1.25}\smash{\begin{tabular}[t]{l}20\end{tabular}}}}%
    \put(0.82050643,0.04342104){\color[rgb]{0.14901961,0.14901961,0.14901961}\makebox(0,0)[lt]{\lineheight{1.25}\smash{\begin{tabular}[t]{l}30\end{tabular}}}}%
    \put(0.85369945,0.04342104){\color[rgb]{0.14901961,0.14901961,0.14901961}\makebox(0,0)[lt]{\lineheight{1.25}\smash{\begin{tabular}[t]{l}40\end{tabular}}}}%
    \put(0.88689232,0.04342104){\color[rgb]{0.14901961,0.14901961,0.14901961}\makebox(0,0)[lt]{\lineheight{1.25}\smash{\begin{tabular}[t]{l}50\end{tabular}}}}%
    \put(0.92008519,0.04342104){\color[rgb]{0.14901961,0.14901961,0.14901961}\makebox(0,0)[lt]{\lineheight{1.25}\smash{\begin{tabular}[t]{l}60\end{tabular}}}}%
    \put(0.95327806,0.04342104){\color[rgb]{0.14901961,0.14901961,0.14901961}\makebox(0,0)[lt]{\lineheight{1.25}\smash{\begin{tabular}[t]{l}70\end{tabular}}}}%
    \put(0.986471,0.04342104){\color[rgb]{0.14901961,0.14901961,0.14901961}\makebox(0,0)[lt]{\lineheight{1.25}\smash{\begin{tabular}[t]{l}80\end{tabular}}}}%
    \put(0.84540875,0.02766397){\color[rgb]{0.14901961,0.14901961,0.14901961}\makebox(0,0)[lt]{\lineheight{1.25}\smash{\begin{tabular}[t]{l}time[s]\end{tabular}}}}%
    \put(0,0){\includegraphics[width=\unitlength,page=20]{joint.pdf}}%
    \put(0.7069775,0.05548043){\color[rgb]{0.14901961,0.14901961,0.14901961}\makebox(0,0)[lt]{\lineheight{1.25}\smash{\begin{tabular}[t]{l}0\end{tabular}}}}%
    \put(0.7069775,0.07281321){\color[rgb]{0.14901961,0.14901961,0.14901961}\makebox(0,0)[lt]{\lineheight{1.25}\smash{\begin{tabular}[t]{l}5\end{tabular}}}}%
    \put(0.69941411,0.09014598){\color[rgb]{0.14901961,0.14901961,0.14901961}\makebox(0,0)[lt]{\lineheight{1.25}\smash{\begin{tabular}[t]{l}10\end{tabular}}}}%
    \put(0.6916406,0.0620984){\color[rgb]{0.14901961,0.14901961,0.14901961}\rotatebox{90}{\makebox(0,0)[lt]{\lineheight{1.25}\smash{\begin{tabular}[t]{l}$f_4$[N]\end{tabular}}}}}%
    \put(0,0){\includegraphics[width=\unitlength,page=21]{joint.pdf}}%
    \put(0.47338252,0.0018489){\color[rgb]{0.14901961,0.14901961,0.14901961}\makebox(0,0)[lt]{\lineheight{1.25}\smash{\begin{tabular}[t]{l}(b) Orientation\end{tabular}}}}%
    \put(0.83274347,0.0025651){\color[rgb]{0.14901961,0.14901961,0.14901961}\makebox(0,0)[lt]{\lineheight{1.25}\smash{\begin{tabular}[t]{l}(c) Inputs\end{tabular}}}}%
  \end{picture}%
\endgroup%